\documentclass{aa}

\usepackage{graphicx}
\usepackage{txfonts}
\usepackage{subcaption}  
\usepackage{hyperref}

\begin{document}

   \title{Determining the statistical significance of meteorite--asteroid pairs using geocentric parameters}

    \author{P.M. Shober \inst{1}}
    
    \institute{
    Laboratoire Temps Espace, CNRS, Observatoire de Paris, PSL Universit\'e, Sorbonne Universit\'e, Universit\'e de Lille 1, UMR 8028 du CNRS, 77 av. Denfert-Rochereau 75014 Paris, France}

   \date{Received June 7, 2025; accepted }
 
  \abstract
   {Orbital similarity between precisely observed meteorite falls and near-Earth asteroids (NEAs) has been presented for decades as evidence that some meteorites are coming directly from these asteroids. However, analysis of the statistical significance of these pairings is mixed. Based on osculating orbital elements, there is no evidence of statistically significant clustering; however, some analyses that account for secular perturbations suggest that streams are present.}
   {We tested the statistical significance of meteorite-dropping fireballs and NEA clustering using the $D_{N}$ similarity function based on four geocentric quantities ($U$, $\theta$, $\phi$, and $\lambda_\odot$).}
   {We calculated the cumulative similarity found between 46 meteorite falls, 535 potential meteorite-dropping fireballs, and 20\,516 NEAs maintained by NEODyS-2, along with 34\,836 NEAs maintained by NASA/JPL HORIZONS. Statistical significance was estimated either by (1) using a kernel density estimation-based method to estimate the sporadic background distribution and thus draw random samples or (2) applying a uniform random solar longitude ($\lambda_\odot$). Each comparison to the synthetic sporadic population was repeated to estimate the $3\sigma$ region for which the cumulative similarity distribution is consistent with random association levels.}
   {The observed $D_{N}$ cumulative similarity distribution of 46 instrumentally observed meteorite falls, 535 potential meteorite-dropping fireballs, and over 30k NEA radiants (estimated using six different radiant methods) reveals no statistically significant excess of similarity between the populations consistent with streams.} 
   {Based on nearly 600 fireball observations and geocentric impact parameters, we find there is no statistically significant clustering between meteorite falls and NEAs. If some meteorites arrive in streams, they make up less than $\sim$\,0.1\% of all falls. Recent asteroid or meteoroid physical processes could still explain features found in meteorites, but this activity is not producing distinguishable orbital streams or pairs.}


   \maketitle

\section{Introduction}
Meteor shower identification has relied heavily on `orbital similarity' functions over the previous half century. These functions enable us to compare pairs of orbits and identify streams that stand out against the sporadic background population. Initially introduced by \citet{southworth1963statistics}, the $D_{SH}$ value is derived by comparing two sets of orbital elements. This measure, which increases as the similarity decreases, is often termed a `dissimilarity parameter'. Several adaptations have been introduced, each slightly changing the degree of influence of different elements on the $D$ value \citep{drummond1981test,steel1991structure,jopek1993remarks,jopek2008meteoroid,jenniskens2008meteoroid}. Although $D$ criteria and other similarity metrics are widely employed, the results can be easily misinterpreted as the significance depends strongly on sample size, relative background flux, and the clustering algorithm chosen (e.g. single-linkage, DBSCAN, etc.; \citealt{jopek1997stream,pauls2005decoherence,koten2014search,egal2017challenge,jopek2017probability,vida2018modelling,shober2024generalizable,jopek2024search,shober2025decoherence}). This is also heightened by the recent recognition that uncertainties in meteor- and fireball-based orbital elements are less precise than once believed \citep{egal2017challenge,vida2018modelling,shober2023comparison}. Several alternative similarity metrics and shower identification methods have also been introduced to overcome these shortcomings \citep{valsecchi1999meteoroid,jenniskens2009report,brown2010meteoroid,eloy2024statistical}.

Many parent bodies have been identified\footnote{\url{https://ceresiaumdc.ta3.sk/listofparents}} using similarity metrics in concurrence with complex numerical integrations of meteoroid streams. The parent bodies generally agreed upon within the scientific community are all related to larger, mostly cometary, meteor showers \citep{jenniskens2006meteor,durisova2024parent}. None of these well-established meteor showers have produced macroscopic meteorite samples, but some studies have explored this possibility \citep{brown2013meteorites}. There have also been a plethora of studies examining the hypothesis that larger, meteorite-dropping fireballs could be associated with streams of material. Most notably, the proposed meteorite pairs of Innisfree--Ridgedale \citep{halliday1987detection}, Příbram--Neuschwanstein \citep{spurny2003photographic}, and Chelyabinsk--NEA 1999 NC43 \citep{borovivcka2013trajectory} have all since been shown to be very likely not statistically significant pairings based solely on orbital similarity \citep{pauls2005decoherence,koten2014search,reddy2015link,shober2025decoherence}. Several works have even proposed that a large portion of meteorites (up to over 50\%) can be directly linked to near-Earth asteroids (NEAs) based on their orbits alone (e.g. \citealp{pena2022orbital,hlobik2024orbital}). However, in these studies, either too large $D$ criteria were used or some other implicit assumptions caused the estimated statistical significance to be unreasonably high \citep{shober2025decoherence}. Clustering within the NEA population has been identified \citep{jopek2020orbital,granvik2024tidal,shober2025decoherence}; however, no considerable statistical evidence to date supports associations between fireballs and such NEAs. 

Nevertheless, the idea that NEAs could be the immediate precursor bodies of at least some meteorites on Earth and form debris streams is not unfounded \citep{borovivcka2015small}. Recent work has identified statistically significant clustering amongst NEAs, likely linked to tidal disruptions during close encounters with the Earth \citep{jopek2020orbital,granvik2024tidal,shober2025decoherence}. \citet{defuentesmarcos2016far} also found that the distributions of $\omega$ and $\Omega$ deviate strongly from uniformity, with some groupings suggested to be due to streams, but most of the deviation observed is produced by secular and Kozai resonances rather than streams. Additionally, analysis of the meteoritic samples themselves, along with observations of recent asteroid activity, has bolstered recent interest in this concept \citep{lauretta2019episodes,turner2021carbonaceous,scott2021short,shober_carbonaceous,jenniskens2025review}. Thus, searching for immediate precursor bodies of meteorite falls amongst NEAs is worthwhile. 

A recent re-analysis of the `decoherence lifetimes' of near-Earth meteoroid streams and the statistical significance of clustering amongst meteorite-dropping fireballs and NEAs found no evidence to support the dozens of claims made over the previous two decades that meteorites can be directly linked to NEAs using their orbits \citep{shober2025decoherence}. However, this study only examined three separate orbital similarity functions \citep{southworth1963statistics,drummond1981test,jopek1993remarks} using the osculating orbital elements at the observation epoch of the fall. 

The dynamical `decoherence' of meteoroid streams (i.e. the time that it takes for a stream to become indistinguishable from the sporadic background) is usually around $\sim$10–50\,kyr for minor streams near the Earth \citep{pauls2005decoherence,shober2025decoherence}. However, during this period, the secular precession of the argument of perihelion ($\omega$) cycles an NEA through a sequence of discrete Earth-intersection geometries. A single orbit can cross the Earth four times per $\omega$ cycle (quadruple crossers) or up to eight times for highly inclined cases, generating multiple distinct geocentric radiants \citep{gronchi2001proper,babadzhanov2012near,pokorny2013opik}. Any fragments ejected within the past few tens of millennia could impact Earth at radiants that bear little resemblance to the present-day orbit of the immediate-precursor body. The practical consequence is that searches confined to a meteorite’s current radiant risk missing earlier intersections elsewhere in the cycle. This mechanism underpins the well-studied 96P/Machholz complex and other stream complexes \citep{babadzhanov2001search,babadzhanov2008meteor,nesluvsan2013meteor,nesluvsan2014meteor,babadzhanov2015extinct,babadzhanov2017investigation,kholshevnikov2016metrics,nesluvsan2021meteoroid}. 

\citet{carbognani2023identifying} sampled the entire $\omega$ cycle for 16\,227 NEAs, finding
30\,740 possible impact radiants; they determined that 20 of the 38 instrumentally observed meteorites can be paired with specific NEAs. However, their significance assessment relied on a single random trial and no formal confidence limits. Here, we build upon the results of \citet{shober2025decoherence} and extend the analysis of \citet{carbognani2023identifying}. While no statistically significant meteorite-dropping streams are observed at the current epoch, it is necessary to test whether these statistically significant associations can be found when searching all possible NEA radiants during their secular $\omega$ cycle.

\section{Data}
We used data from 46 instrumentally observed recovered meteorite falls along with 535 possible meteorite falls. The 535 possible falls were observed by the Global Fireball Observatory (GFO\footnote{\url{https://gfo.rocks/}}; \citealp{devillepoix2020global}), the European Fireball Network (EFN; \citealp{borovivcka2022_one,borovivcka2022_two}), or the Fireball Recovery and InterPlanetary Observation Network (FRIPON\footnote{\url{https://www.fripon.org/}}; \citealp{colas2020fripon}). The possible falls were identified using the $\alpha$-$\beta$ criterion \citep{sansom2019determining} for GFO and FRIPON observations, with a minimal final mass of at least 1\,g and $\geq$\,20\% atmospheric deceleration. The EFN possible meteorite-dropping subset was identified in \citet{borovivcka2022_one}. For further information about the data, please refer to the Methods section of \citet{shober2025perihelion}, from which these data were taken. The 46 meteorite falls used can be found in \ref{tab:meteorite_falls}. 

\section{Methods}

\subsection{$D_{N}$ criterion} 
Classical $D$ parameters compare two orbits in the five-element heliocentric space $(a,e,i,\Omega$, and $\omega$), but that strategy suffers from two fundamental drawbacks. First, a meteor’s
heliocentric elements are derived from a short fireball arc and are far less precise than those of asteroids. Second, those elements evolve on $10^{3}$–$10^{5}$ yr timescales under planetary
perturbations, so a small $D$ value measured today is not, by itself, proof of common origin \citep{pauls2005decoherence,babadzhanov2012near,pokorny2013opik,babadzhanov2015extinct,babadzhanov2017investigation,shober2025decoherence}. To bypass these limitations, \citet{valsecchi1999meteoroid} introduced the distance function $D_{N}$, defined in a space of four geocentric observables that are directly derived from the observations (unlike orbital elements) and thus matched to the actual dimensionality of a well-observed fireball. The four coordinates are the magnitude of the unperturbed geocentric velocity ($U$), the two Öpik encounter angles ($\theta$ and $\phi$), which specify the direction of $U$ after removing Earth’s gravitational deflection, and the solar longitude at impact ($\lambda_\odot$).

The similarity criterion, \( D_{N} \), is defined as
\begin{align}
    D^{2}_{N} &= (U_2 - U_1)^2 + w_1 (\cos \theta_2 - \cos \theta_1)^2 + \Delta\xi^2 ,\\
    \text{where} \\ 
    \Delta\xi^2 &= \min \left( w_2 \Delta\phi_I^2 + w_3 \Delta \lambda_I^2, \, w_2 \Delta\phi_{2}^2 + w_3 \Delta \lambda_{2}^2 \right) \\
    \Delta\phi_I &= 2 \sin\left(\frac{\phi_2 - \phi_1}{2}\right) \\
    \Delta\phi_{2} &= 2 \sin\left(\frac{180^\circ - \phi_2 - \phi_1}{2}\right) \\
    \Delta \lambda_1 &= 2 \sin\left(\frac{\lambda_2 - \lambda_1}{2}\right) \\
    \Delta \lambda_{2} &= 2 \sin\left(\frac{180^\circ - \lambda_2 - \lambda_1}{2}\right),
\end{align}
and $w_1$ , $w_2$ , and $w_3$ are suitably defined weighting factors. All weightings were set to 1.0 here, as this is standard practice and was originally used in \citealp{jopek1999meteoroid}. Note that $\Delta\xi$ is small if $\phi_1 - \phi_2$ and $\lambda_1 - \lambda_2$ are either both small or both close to $180^{\circ}$.

\subsection{NEA radiant methods} \label{subsec:radiant_methods}

To replicate the analysis of \citet{carbognani2023identifying}, this analysis also uses the list of NEA geocentric encounter conditions $U$, $\theta$, $\phi$, and $\lambda_\odot$ maintained by NEODyS-2\footnote{\url{4https://newton.spacedys.com/~neodys2/propneo/encounter.cond}}. This list was computed with the ORBFIT software\footnote{\url{5http://adams.dm.unipi.it/orbfit/}} that gives the encounter conditions for 20\,516 NEAs of the total 86\,946 possible radiants. \citet{carbognani2023identifying} used radiants from only 16\,227 NEAs; otherwise, the methodology of the analysis is identical. 

To confirm that the results are not tied to a specific orbit–to–radiant conversion, we augmented the NEODyS/ORBFIT set with five additional theoretical radiant lists generated by the B, A, W, H, Q algorithms in Neslušan’s program \citep{neslusan1998computer}. These routines all force an orbit to cross Earth’s path but differ in which elements are minimally adjusted: (i) in the Q (perihelion-shift) method, only $q$ is varied \citep{hasegawa1990predictions}; (ii) the B method shifts both $q$ and $e$ \citep{svoren1993applicability}; (iii) the W method rotates the line of apsides ($\omega$ shift) \citep{steel1985collisions}; (iv) the A method rotates the orbit about the apsidal line \citep{svoren1993applicability}; and (v) the H method simultaneously tweaks $\omega$ and $i$ \citep{hasegawa1990predictions}.  

Radiants were computed for all 34\,836 NEAs whose osculating elements are archived in the Jet Propulsion Laboratory (JPL HORIZONS system\footnote{\url{https://ssd.jpl.nasa.gov/horizons}} -- a dataset more than twice the size used by \citet{carbognani2023identifying}. By analysing the $D_{N}$ statistics derived from the six independent radiant sets, we can test whether any meteorite–NEA pairing is robust to both the choice of radiant algorithm and the inclusion of a more complete NEA catalogue.

\subsection{Estimating sporadic distributions}\label{subsec:sporadic}
Two methods were used to estimate the sporadic background population and test the statistical significance of observed clustering. To replicate the \citet{carbognani2023identifying} methodology, we kept the NEODyS/ORBFIT parameters $(U,\theta$, and $\phi$) of every NEA and meteorite-dropping fireball fixed but replaced $\lambda_\odot$ with a value drawn from a uniform distribution between 0$^{\circ}$ and 360$^{\circ}$. This effectively randomises the date of the encounter with the Earth. This randomisation must be done for both NEA radiant and fireball populations, as the null hypothesis under consideration is that both fireball and NEA geocentric parameters are sporadic. \citet{carbognani2023identifying} only randomised the $\lambda_\odot$ for the NEA population; however, this implicitly assumes that the meteorite geocentric parameters are not clustered. If this is not true, i.e. there are streams, any real streams within the observed dataset could have their statistical significance mischaracterised because they were assumed to be sporadic in the null test. A similar method was used by \citet{pauls2005decoherence}, except they randomised the $\Omega$ of the samples. 

\begin{figure*}
    \centering
    \includegraphics[width=\linewidth]{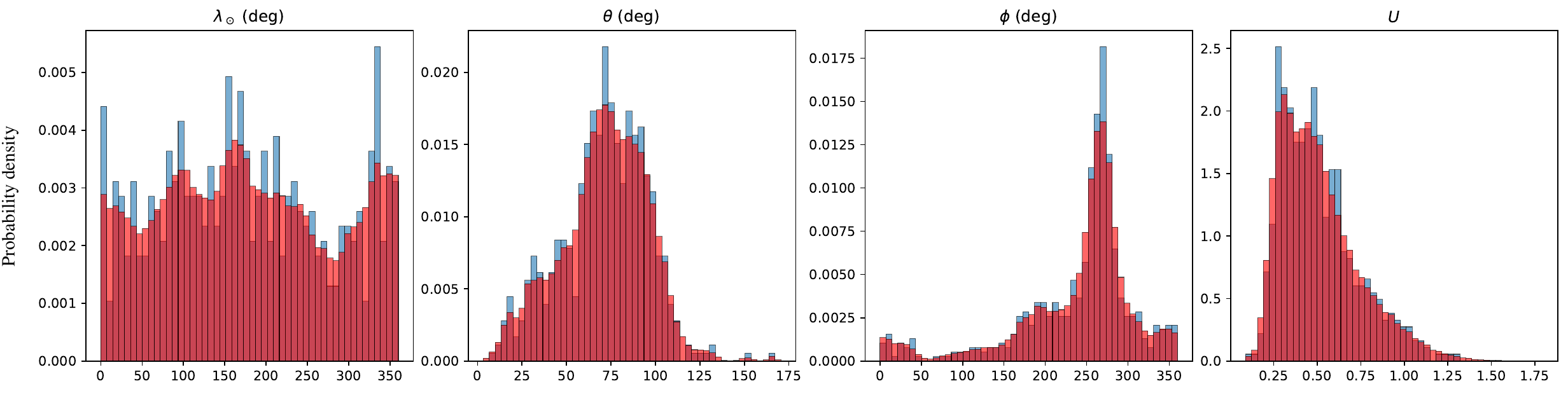}
    \caption{Geocentric parameter distribution of 535 potential meteorite-dropping fireballs (blue) versus the KDE-estimated PDF (red) using a Gaussian kernel with a bandwidth of [0.2564, 0.2564, 0.2489, 0.2564, 0.1619, 0.2333, 0.2564]. This PDF, which approximates the corresponding observed sporadic population, was used to estimate the degree of random association in the population through Monte Carlo simulations.}
    \label{fig:535_KDE}
\end{figure*}

\begin{figure*}
    \centering

    \begin{subfigure}{\linewidth}
        \centering
        \includegraphics[height=3.5cm]{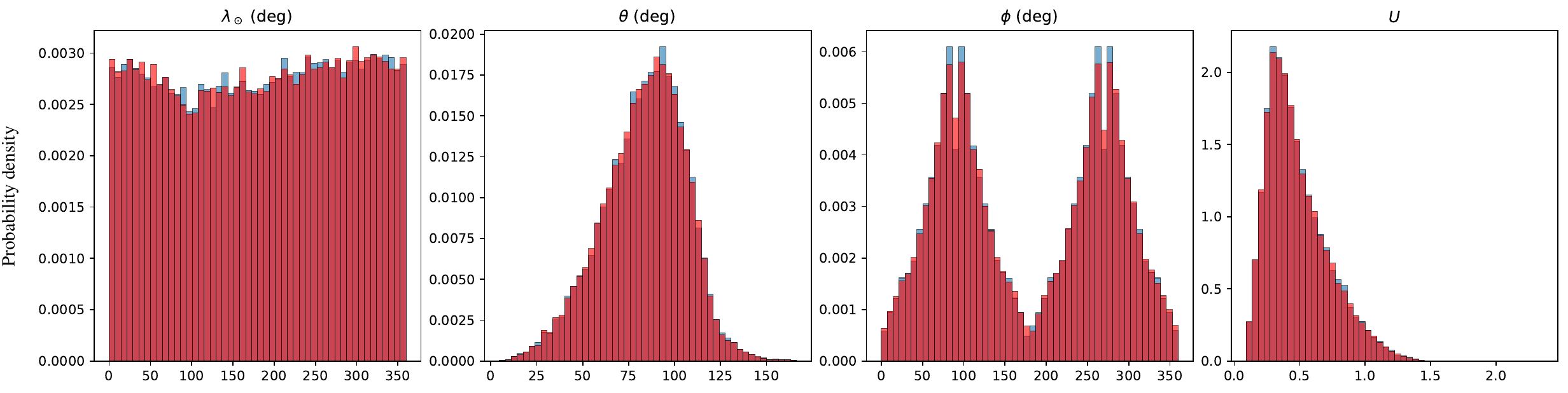}
        \caption{\centering }
        \label{fig:plot1}
    \end{subfigure}

    \begin{subfigure}{\linewidth}
        \centering
        \includegraphics[height=3.5cm]{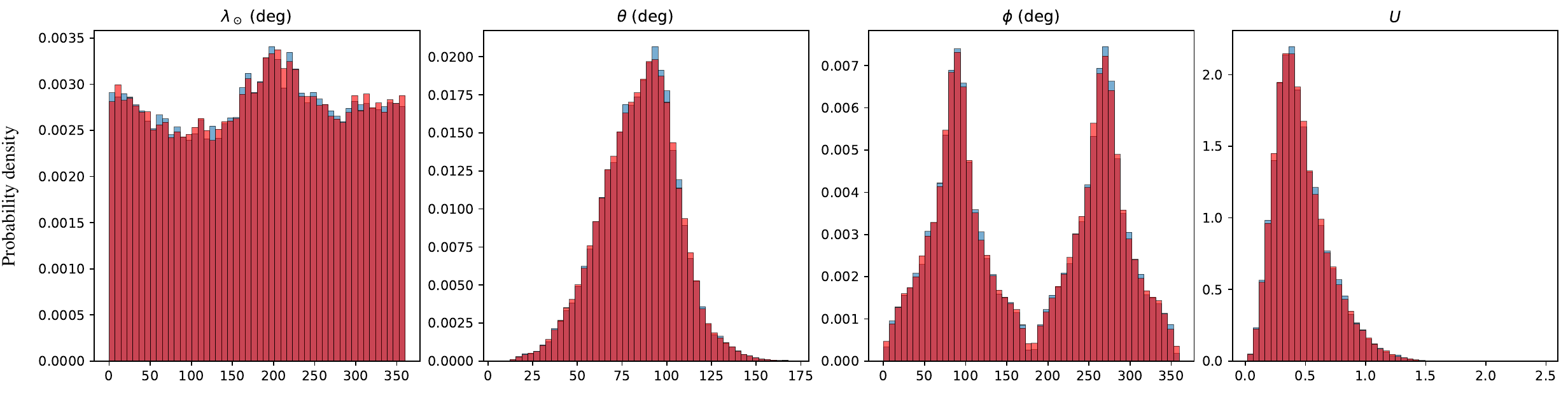}
        \caption{\centering }
        \label{fig:plot2}
    \end{subfigure}

    \begin{subfigure}{\linewidth}
        \centering
        \includegraphics[height=3.5cm]{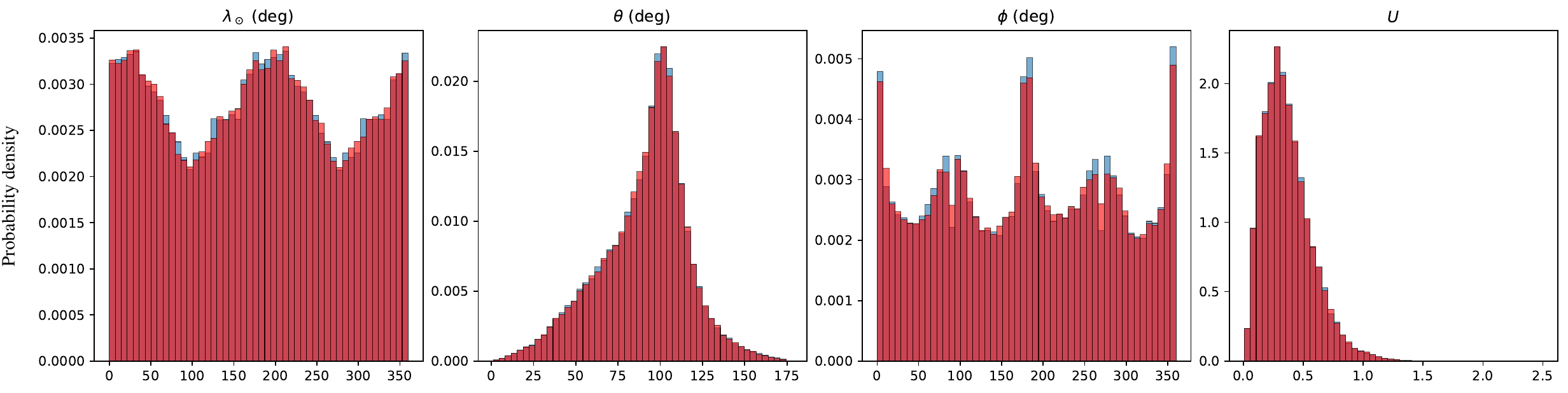}
        \caption{\centering }
        \label{fig:plot3}
    \end{subfigure}

    \begin{subfigure}{\linewidth}
        \centering
        \includegraphics[height=3.5cm]{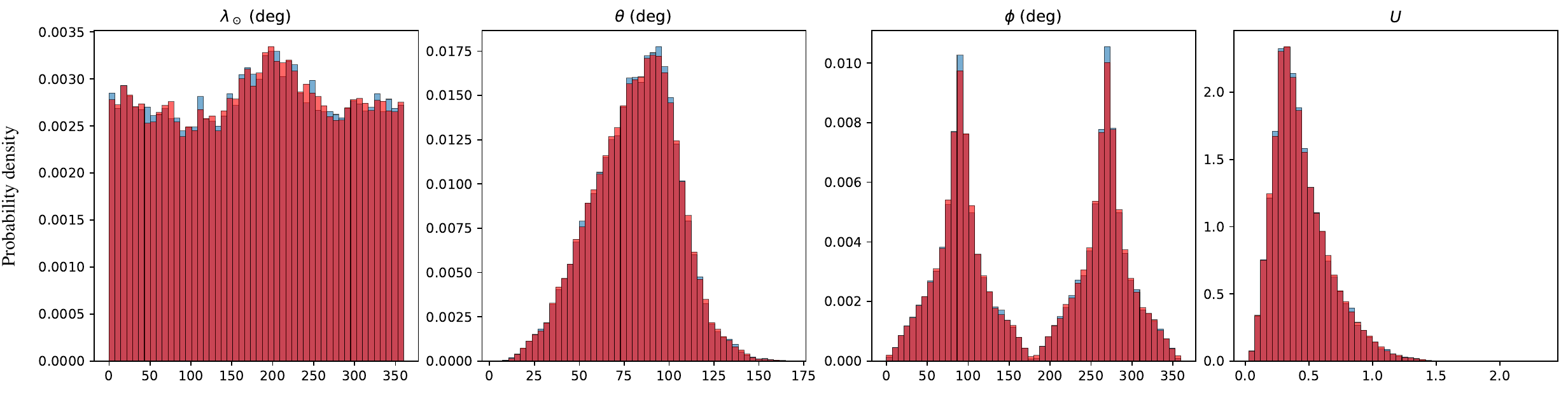}
        \caption{\centering }
        \label{fig:plot4}
    \end{subfigure}

    \begin{subfigure}{\linewidth}
        \centering
        \includegraphics[height=3.5cm]{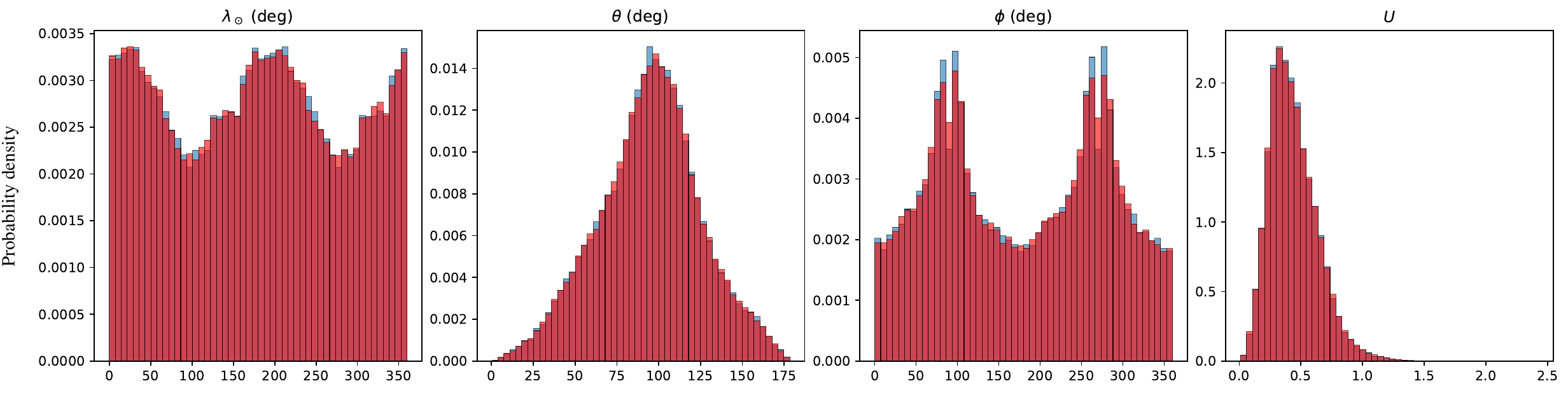}
        \caption{\centering }
        \label{fig:plot5}
    \end{subfigure}

    \begin{subfigure}{\linewidth}
        \centering
        \includegraphics[height=3.5cm]{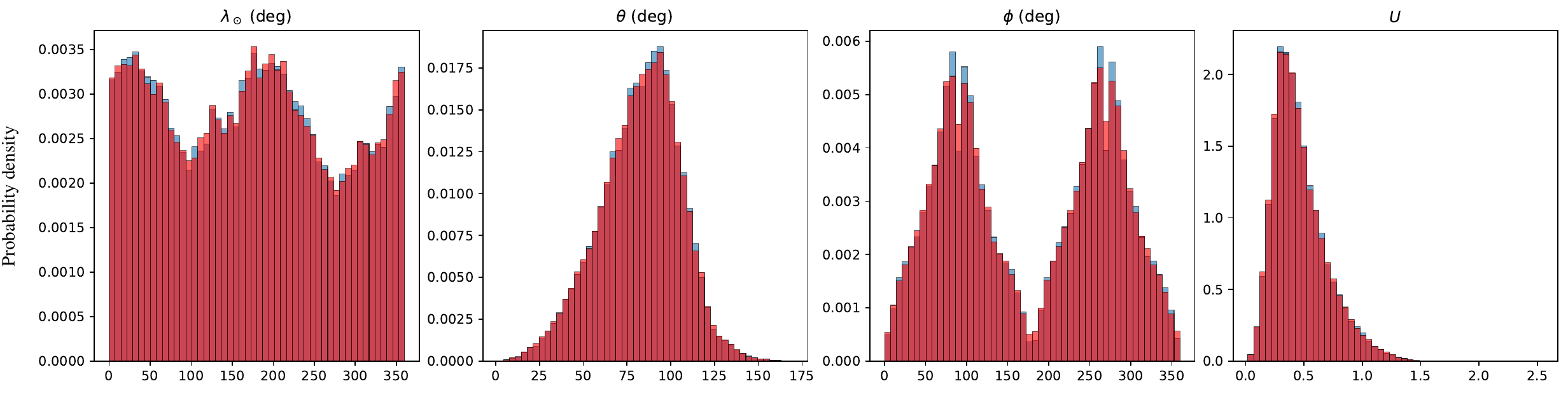}
        \caption{\centering }
        \label{fig:plot6}
    \end{subfigure}

    \caption{Geocentric parameter distribution of estimated NEA radiants (blue) versus the KDE-estimated PDF (red) using a Gaussian kernel: NEODyS-2 (a), Method A (b), Method B (c), Method H (d), Method Q (e),  and Method W (f). These PDFs were used to estimate the degree of random association in the population through Monte Carlo simulations.}
    \label{fig:KDE_all_six}
\end{figure*}

Furthermore, although randomising the orientations, either $\lambda_\odot$ or $\Omega$, has been previously used to estimate sporadic distributions, they assume uniform distributions that are not consistent with actual sporadic observations (see Fig.~\ref{fig:535_KDE} or~\ref{fig:KDE_all_six}). The sporadic distribution of meteoroids and asteroids do not have uniform $\lambda_\odot$ or $\Omega$ distributions \citep{jeongahn2014non}, and the observed populations have observational biases that further skew the distributions \citep{halliday1996detailed,granvik2018debiased}. Seasonal effects, resonances, and observational biases change and modify these distributions. If the biases are not present, the expected degree of similarity between the populations will drop precipitously. For example, in \citet{hlobik2024orbital}, they drew random samples from the de-biased near-Earth object (NEO)  model of \citet{granvik2018debiased} and compared them to the orbits of recovered meteorites. This is fatally flawed, as the populations being compared are not the same as the populations being tested. The distribution of known NEOs exhibits significant observational biases, resulting in a greater overall similarity within the population compared to the terrestrial impact populations. Thus, it is critical to test the degree of similarity expected for the sporadic observed populations. Otherwise, deviations in the amount of similarity could appear that are unrelated to the problem of statistically significant meteorite-NEA streams. 

We also modelled the sporadic background for both the meteorite-fall and NEA-radiant populations with kernel density estimation (KDE; \citealt{vida2017generating,shober2024generalizable}). The probability density function (PDF) output based on the 535 possible meteorite falls was also used to generate random samples for the 46 recovered meteorites. A KDE places a smooth kernel, here a multivariate Gaussian, on every data point and sums those kernels to obtain a continuous estimate of the underlying PDF. Unlike a histogram, KDE is non-parametric, free of arbitrary bin edges, and is one of the most effective methods for generating realistic synthetic samples \citep{vida2017generating,shober2024generalizable}. One can draw arbitrarily large Monte Carlo samples from it, obtaining synthetic radiants that preserve the full covariance structure of the original data while seamlessly smoothing out any remaining minor clusters, precisely the behavior desired for significance testing. However, it is essential to note that while this approach is seen as an improvement over simply randomising the impact date, it is extremely easy to obtain unrealistic results if the parameters are not rigorously inspected to be realistic (i.e. within the correct range, maintaining the cyclic nature of angles, etc.). In this study, after Z-score standardisation of the data, we transformed the three cyclic angles $(\lambda_\odot,\theta$, and $\phi$) to Cartesian pairs $(\cos\alpha$ and $\sin\alpha)$, ensuring circular continuity and the full covariance structure of the data in the seven-dimensional space $(\cos\lambda_\odot,\sin\lambda_\odot,\cos\theta,\sin\theta,\cos\phi,\sin\phi,$ and $U)$. A diagonal, anisotropic Gaussian kernel was fitted with bandwidths $h_i$ obtained from the improved Sheather–Jones (ISJ) estimator (\citealp{sheather1991reliable,botev2010kernel}); this automatically balances over- and under-smoothing across dimensions while retaining inter-parameter correlations. The mean integrated squared error (MISE) gauges the accuracy of a kernel-density estimate $\hat{f}(x;h)$,
\begin{equation}
    \operatorname{MISE}(h)
    = \mathbb{E}\!\Bigl[\,
           \int\bigl(\hat{f}(x;h)-f(x)\bigr)^{2}\,dx
           \Bigr]
    \label{eq:MISE}
,\end{equation}
where $h$ is the kernel bandwidth. In practical terms, $h$ sets the trade-off between noise and smoothing: narrow kernels follow every bump in the data but amplify noise, whereas wide kernels smooth noise at the cost of erasing real structure. ISJ chooses the bandwidth that minimises the large-sample limit of the asymptotic MISE (AMISE). The AMISE depends on the `roughness' of the true density, $\|f''\|^{2}$, a quantity that cannot be measured directly, so the ISJ algorithm estimates that roughness from the data in a self-consistent, recursive fashion \citep{botev2010kernel}. The bandwidth returned by ISJ is therefore the one that minimises the expected squared difference between the KDE and the true, but unseen, density, yielding a smoothing level that retains genuine orbital structure while filtering out sampling noise. ISJ is ideal for orbital or radiant distributions, as they can be multimodal and generally not normally distributed.

\subsection{Statistical significance}
To assess the statistical significance of any excesses of similarity based on $D_{N}$, a methodology similar to that of \citet{shober2025decoherence} was used. Monte Carlo experiments were conducted where $N_{population}$ draws, equal in size to the original datasets, were randomly generated from our sporadic background distribution estimates (as described in Sect.~\ref{subsec:sporadic}). The $D_{N}$ value was then calculated for every possible pair combination between the sporadic meteorite-dropping fireball population and sporadic NEA radiant population. The cumulative similarity distribution (CSD), which describes the total number of pairs that are $<D_{N}$ as a function of $D_{N}$, was estimated for the 46 recovered meteorites and each of the six radiant methods described in Sect.~\ref{subsec:radiant_methods}. Additionally, the CSDs of the 535 possible falls and the six different NEA radiant distributions were also calculated. In total, we tested the statistical significance of 12 different CSDs to determine whether a statistically significant amount of similarity exists among the 46 recovered falls or the 535 fireballs, based on geocentric parameters. This was then carried out ten times for each CSD to estimate the $3\sigma$ confidence region, which describes the amount of similarity expected within the population due to random associations. If an ongoing mechanism were truly creating meteoroid streams in near-Earth space, the distribution of $D_{N}$ values would show a marked surplus at the low end of the CSD distribution, a deviation from a power law \citep{vokrouhlicky2008pairs,rozek2011orbital,shober2025decoherence}. Using synthetic sporadics generated with the randomised $\lambda_\odot$ or the KDE-based Monte Carlo tests, we evaluated $D_{N}$ for more than 7$\times10^{8}$ unique combinations.

\section{Results}

\begin{figure*}
    \centering

    \begin{subfigure}{0.48\linewidth}
        \centering
        \includegraphics[width=\linewidth]{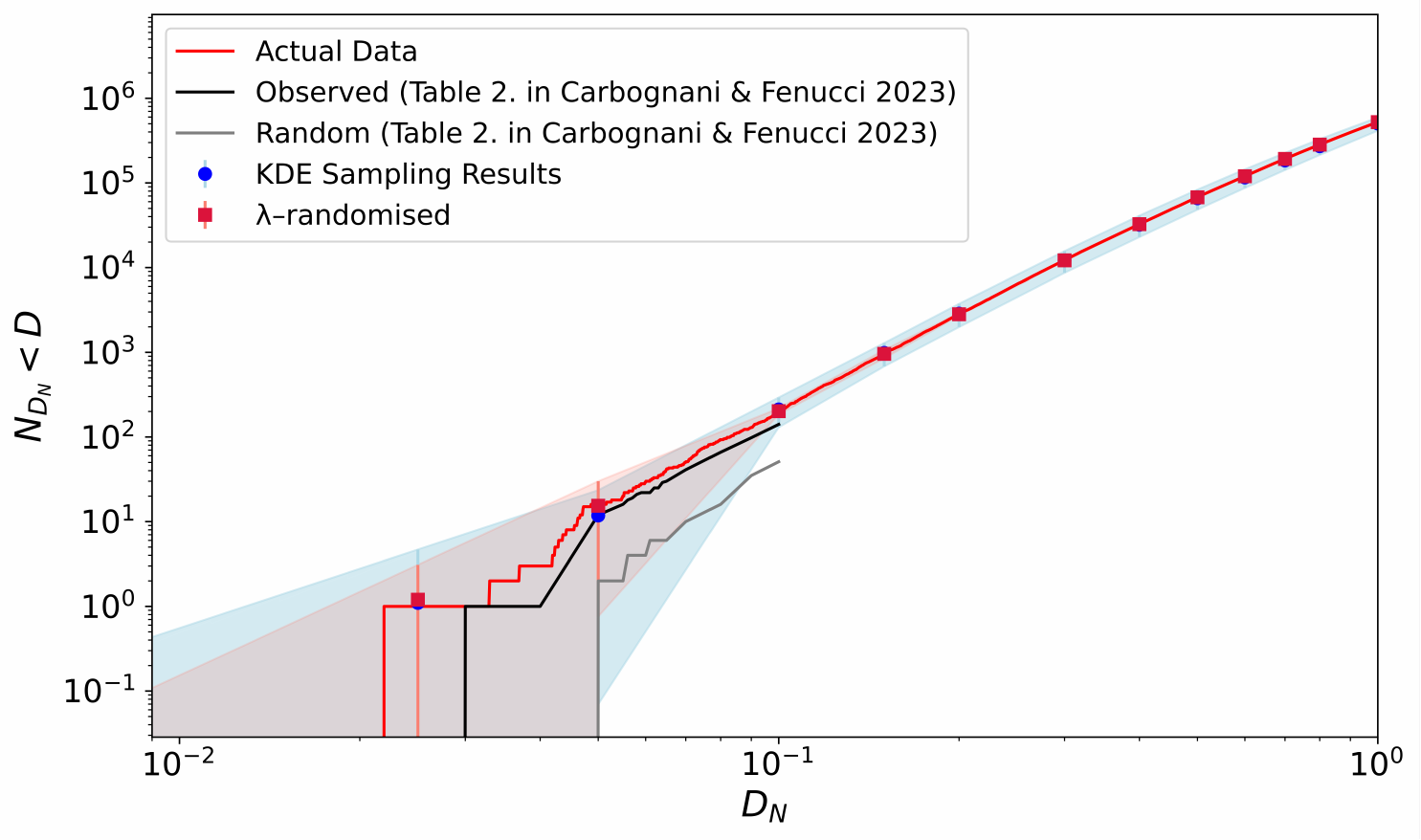}
        \caption{\centering  }
        \label{fig:result1}
    \end{subfigure}
    \hfill
    \begin{subfigure}{0.48\linewidth}
        \centering
        \includegraphics[width=\linewidth]{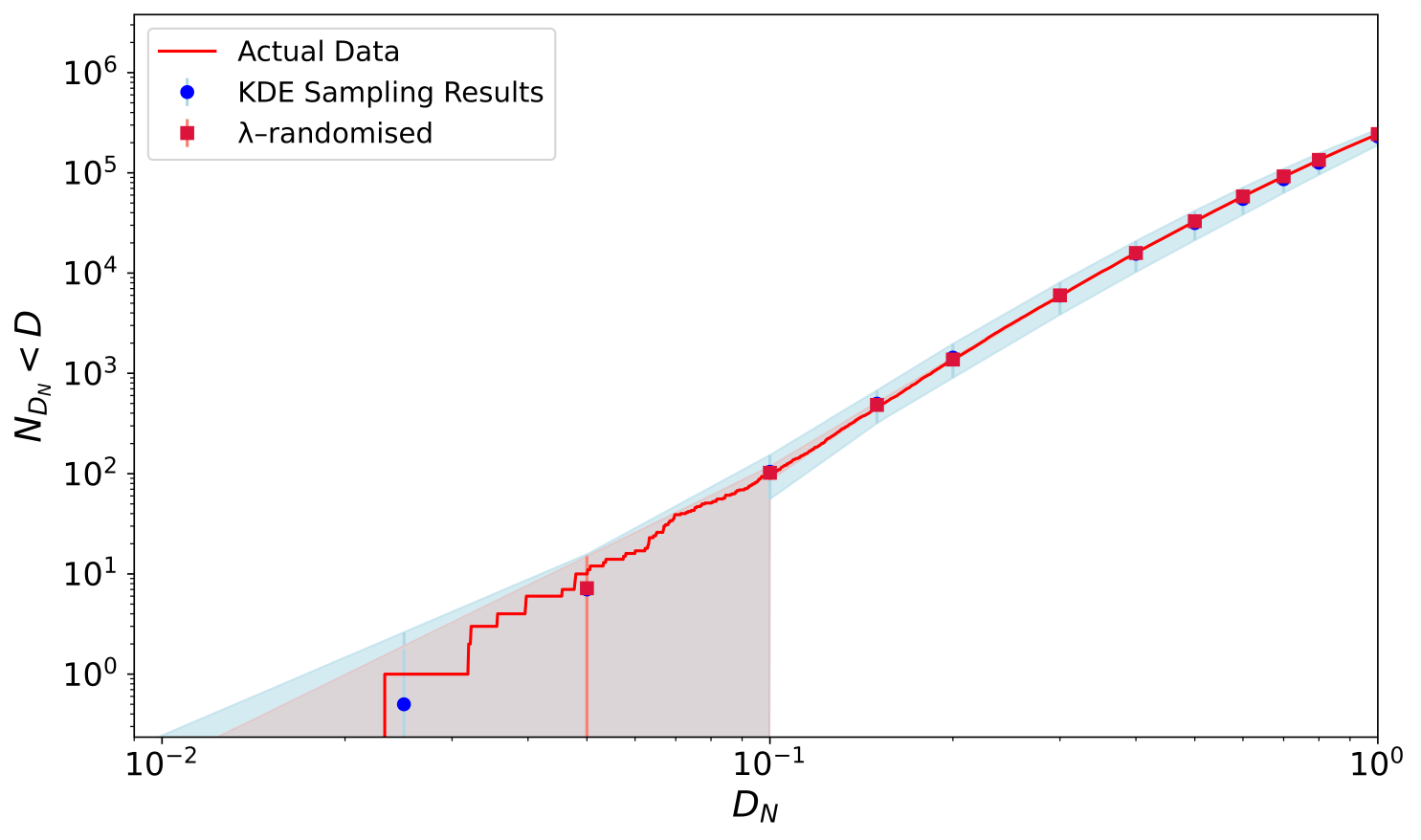}
        \caption{\centering }
        \label{fig:result2}
    \end{subfigure}

    \vspace{0.4cm}

    \begin{subfigure}{0.48\linewidth}
        \centering
        \includegraphics[width=\linewidth]{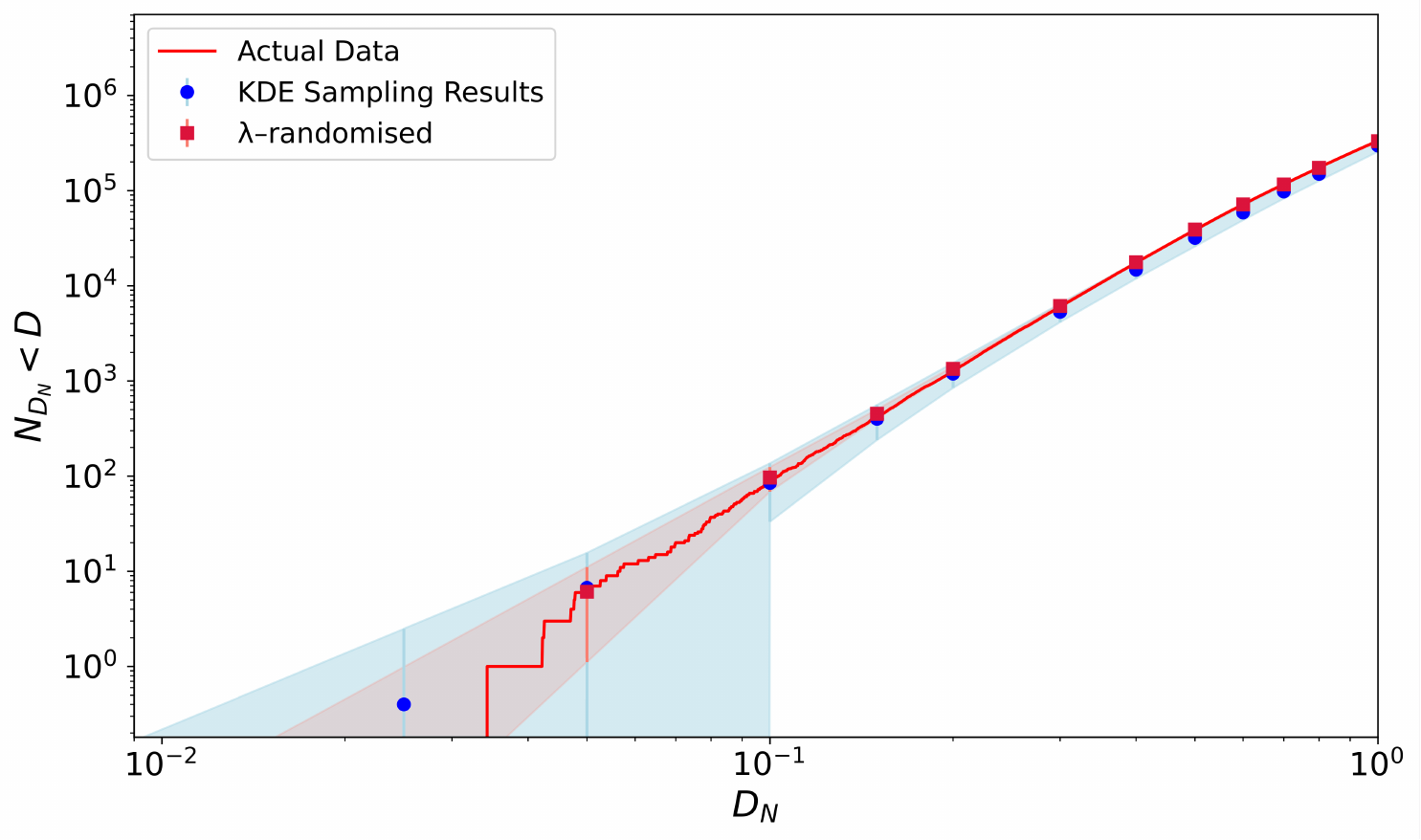}
        \caption{\centering  }
        \label{fig:result3}
    \end{subfigure}
    \hfill
    \begin{subfigure}{0.48\linewidth}
        \centering
        \includegraphics[width=\linewidth]{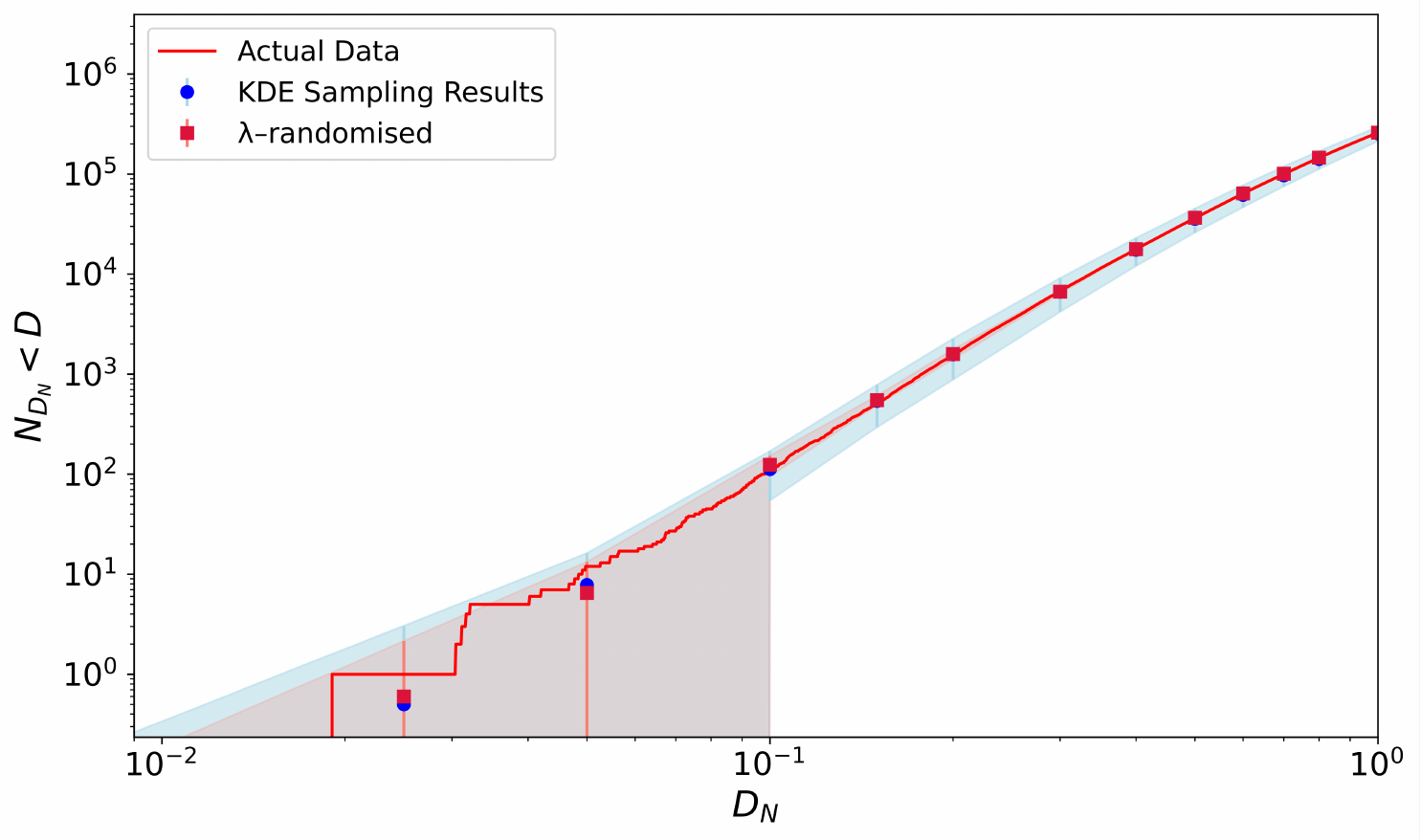}
        \caption{\centering  }
        \label{fig:result4}
    \end{subfigure}

    \vspace{0.4cm}

    \begin{subfigure}{0.48\linewidth}
        \centering
        \includegraphics[width=\linewidth]{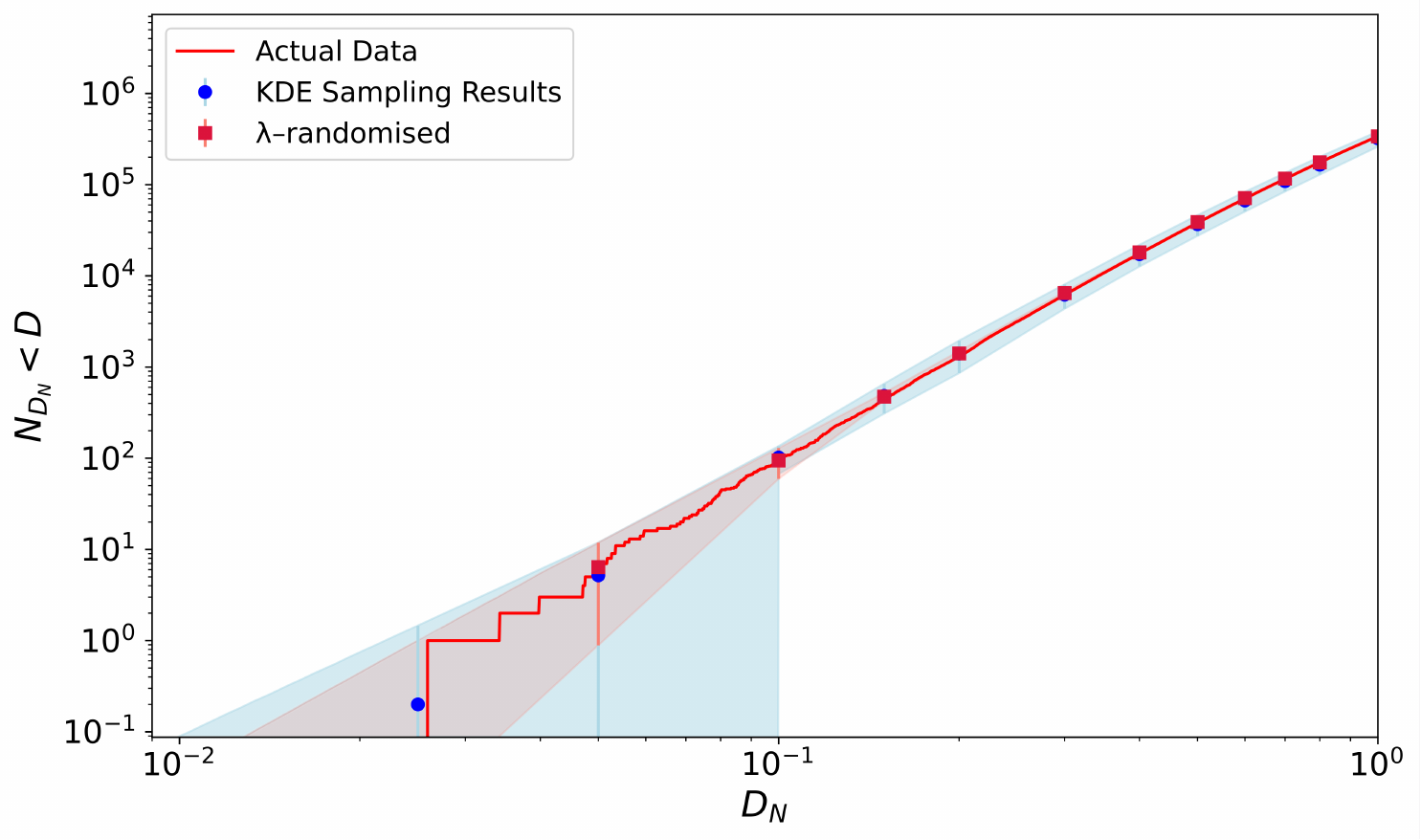}
        \caption{\centering  }
        \label{fig:result5}
    \end{subfigure}
    \hfill
    \begin{subfigure}{0.48\linewidth}
        \centering
        \includegraphics[width=\linewidth]{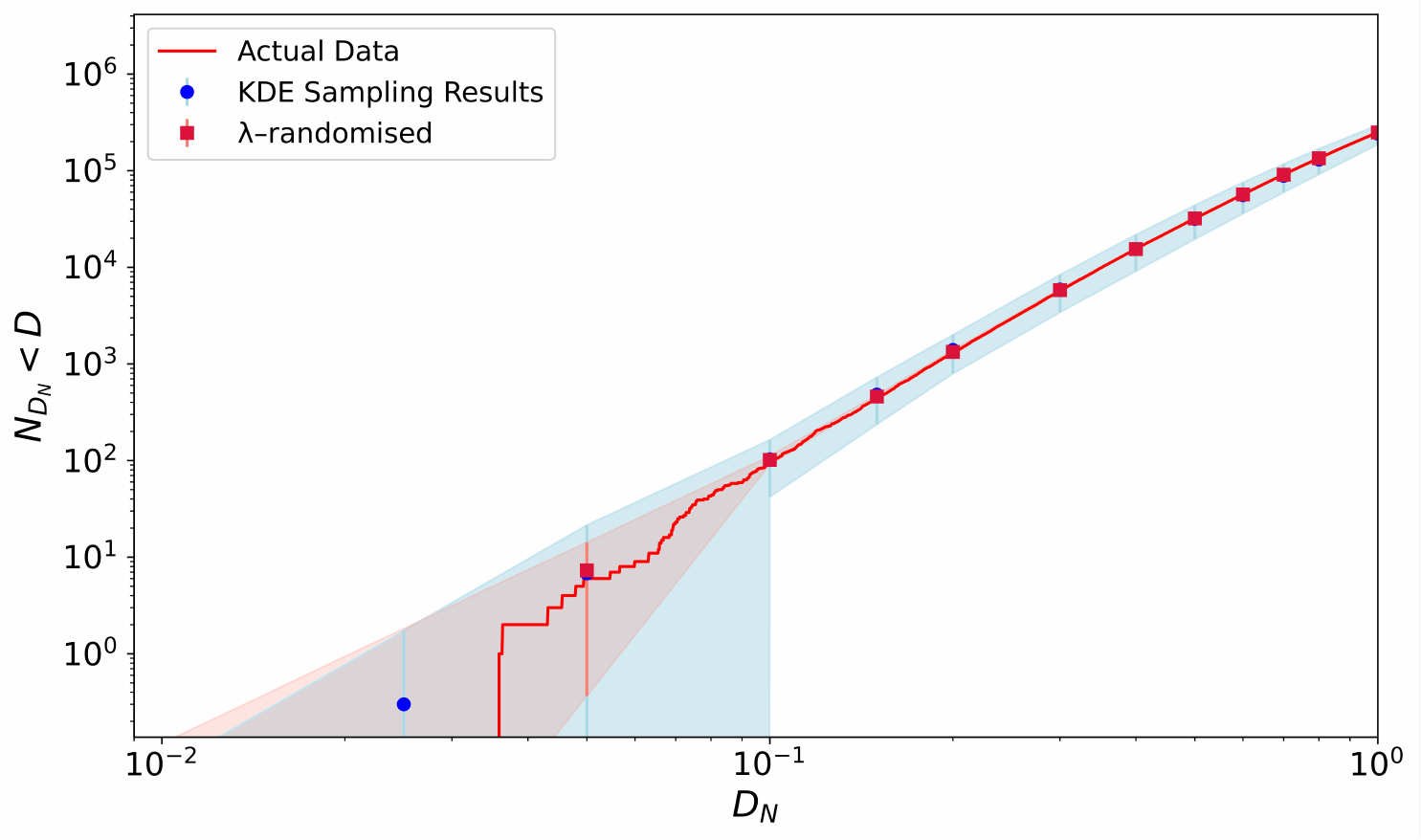}
        \caption{\centering  }
        \label{fig:result6}
    \end{subfigure}

    \caption{CSDs of the 46 recovered meteorite falls and NEA theoretical impact radiants: NEODyS-2 (a), Method A (b), Method B (c), Method H (d), Method Q (e), and Method W (f).     The red curve, labelled `Actual Data', represents the cumulative similarity between every possible radiant--fall pair combination for the 46 falls and the theoretical NEA radiants of a particular method. The NEA radiants were estimated using six different methods, one for each panel. Two sporadic association regions are estimated. The blue region encompasses the $3\sigma$ confidence region for random associations based on the KDE-drawn samples. The red region represents the $3\sigma$ confidence region for random associations based on a uniformly random $\lambda_\odot$. In panel (a), the `observed' and `random' curves reported by \citet{carbognani2023identifying} are also included for comparison purposes.}
    \label{fig:46_CSDs}
\end{figure*}

\subsection{Kernel density estimates of the sporadic background}
\label{subsec:kde_results}

A multivariate Gaussian KDE was fitted to each radiant sample (Fig.~\ref{fig:535_KDE} for the 535 potential falls;  Fig.~\ref{fig:KDE_all_six} for the six NEA catalogues). Figure~\ref{fig:535_KDE} shows that the resulting meteorite fall PDF reproduces the seasonal modulation in $\lambda_\odot$ produced mainly by weather and radiant-visibility biases. Non-negligible non-uniformities appear in the NEA radiants as well (Fig.~\ref{fig:KDE_all_six}), which are expected and consistent with the results of \citet{jeongahn2014non}. These red regions are the PDFs from which all Monte Carlo `sporadic' samples were drawn. For comparison, we also generated control samples by holding $(U,\theta,\phi)$ fixed and assigning a uniformly random $\lambda_\odot$, as done in \citet{carbognani2023identifying}.

Method-dependent differences are evident in the fitted radiant PDFs (Fig.~\ref{fig:KDE_all_six}), reflecting how each algorithm enforces Earth intersection. In particular, Method A (rotation about the apsidal line; \citealt{svoren1993applicability}) reorients the orbital plane by changing $(i,\Omega)$ while leaving $(q,e,\omega)$ fixed, which drives the asymptotic geocentric approach azimuth to cluster at $\phi \simeq 0^\circ$ and $180^\circ$. In contrast, the Q/B/W/H variants modify \(q\), \(e\), or rotate the line of apsides (\(\omega\)) while leaving the plane orientation \((i,\Omega)\) unchanged or only weakly perturbed, thereby shifting the Earth-intersection to different true anomalies within an essentially fixed plane and yielding mutually similar \(\phi\) distributions that lack the \(0^{\circ}/180^{\circ}\) concentration characteristic of Method~A (often with excess near \(90^{\circ}/270^{\circ}\)). The standard-scaled ISJ bandwidths are listed in Table~\ref{tab:bandwidths}.

\begin{table*}[t]
\caption{ISJ bandwidth vectors (standard-score space).}
\label{tab:bandwidths}
\centering
\small
\begin{tabular}{lccccccc}
\hline\hline
Sample & $h_{\!\text{sol\,x}}$ & $h_{\!\theta_x}$ & $h_{\!\phi_x}$ & $h_{\!\text{sol\,y}}$ & $h_{\!\theta_y}$ & $h_{\!\phi_y}$ & $h_U$\\
\hline
535 potential falls      & 0.256 & 0.256 & 0.249 & 0.256 & 0.162 & 0.233 & 0.256\\
46 recovered falls       & 0.423 & 0.402 & 0.423 & 0.423 & 0.209 & 0.423 & 0.299\\
NEODyS/ORBFIT            & 0.0926& 0.0926& 0.0926& 0.0926& 0.0539& 0.0926& 0.0926\\
Method A                 & 0.109 & 0.109 & 0.109 & 0.109 & 0.0653& 0.109 & 0.105\\
Method B                 & 0.0968& 0.0966& 0.0968& 0.0968& 0.0556& 0.0968& 0.0918\\
Method H                 & 0.108 & 0.108 & 0.0996& 0.108 & 0.0723& 0.108 & 0.106\\
Method Q                 & 0.0968& 0.0968& 0.0968& 0.0968& 0.0720& 0.0968& 0.0946\\
Method W                 & 0.108 & 0.108 & 0.108 & 0.108 & 0.0636& 0.108 & 0.106\\
\hline
\end{tabular}
\end{table*}

\subsection{Cumulative similarity statistical significance}
\begin{figure*}
    \centering

    \begin{subfigure}{0.48\linewidth}
        \centering
        \includegraphics[width=\linewidth]{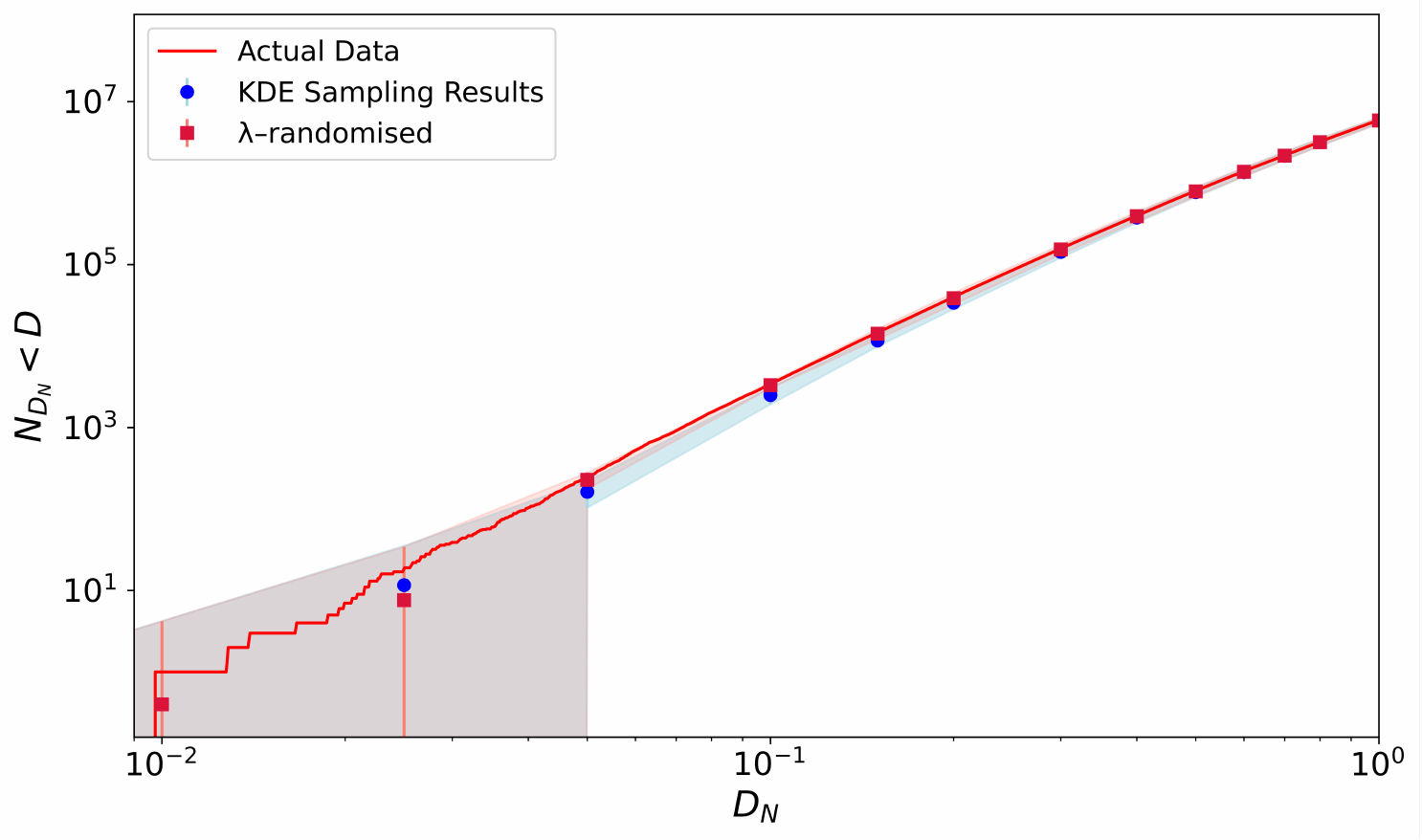}
        \caption{\centering  }
        \label{fig:result1}
    \end{subfigure}
    \hfill
    \begin{subfigure}{0.48\linewidth}
        \centering
        \includegraphics[width=\linewidth]{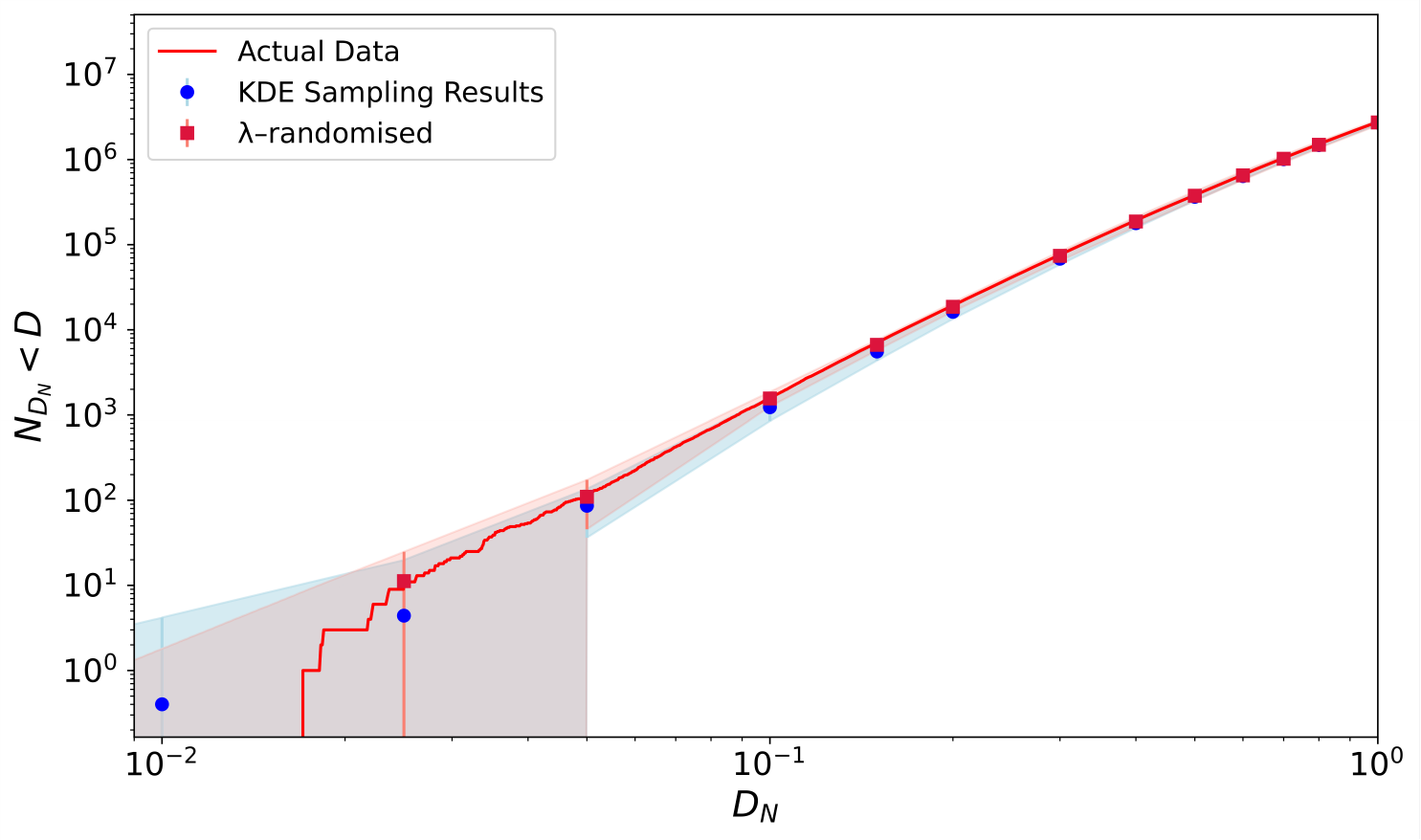}
        \caption{\centering  }
        \label{fig:result2}
    \end{subfigure}

    \vspace{0.4cm}

    \begin{subfigure}{0.48\linewidth}
        \centering
        \includegraphics[width=\linewidth]{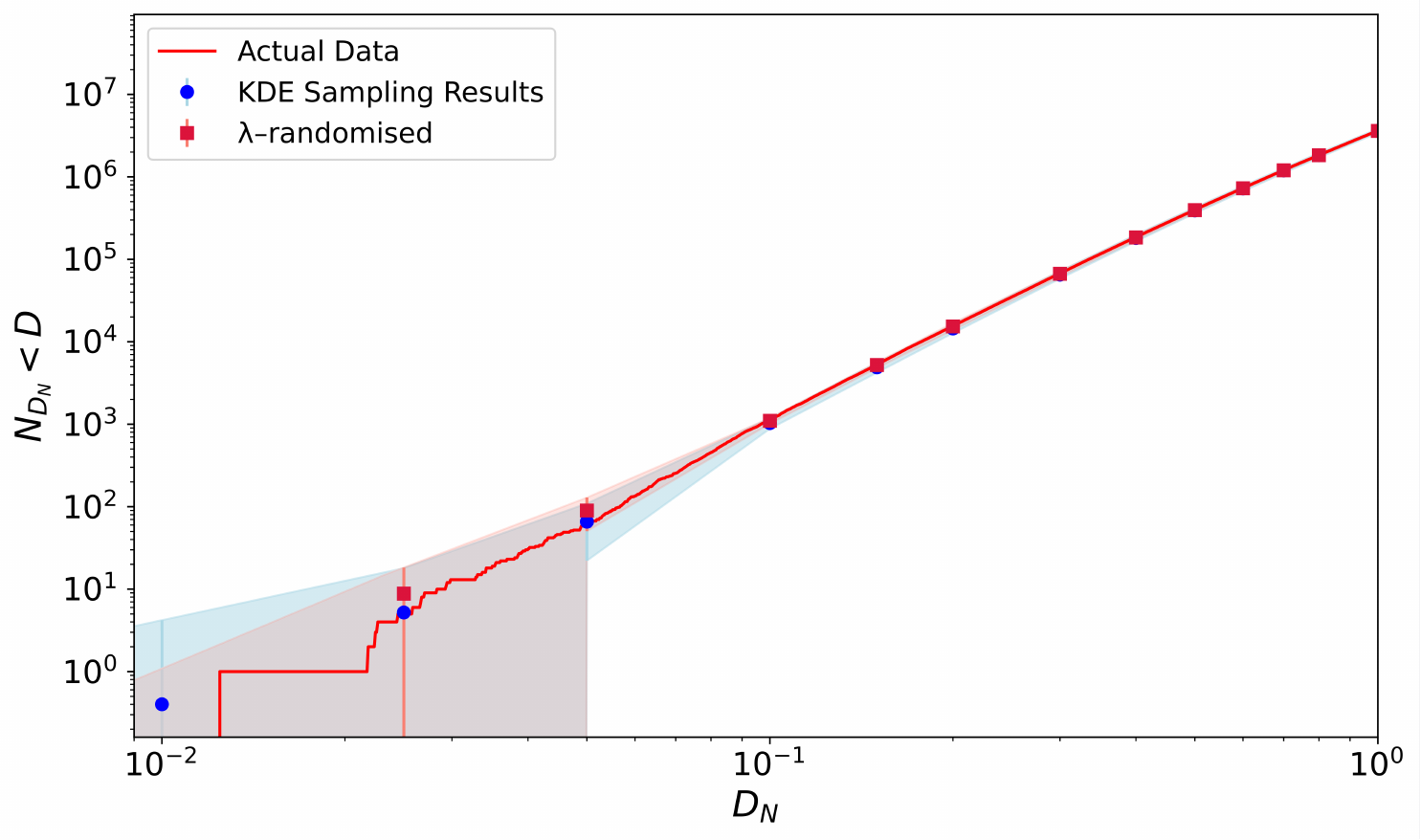}
        \caption{\centering  }
        \label{fig:result3}
    \end{subfigure}
    \hfill
    \begin{subfigure}{0.48\linewidth}
        \centering
        \includegraphics[width=\linewidth]{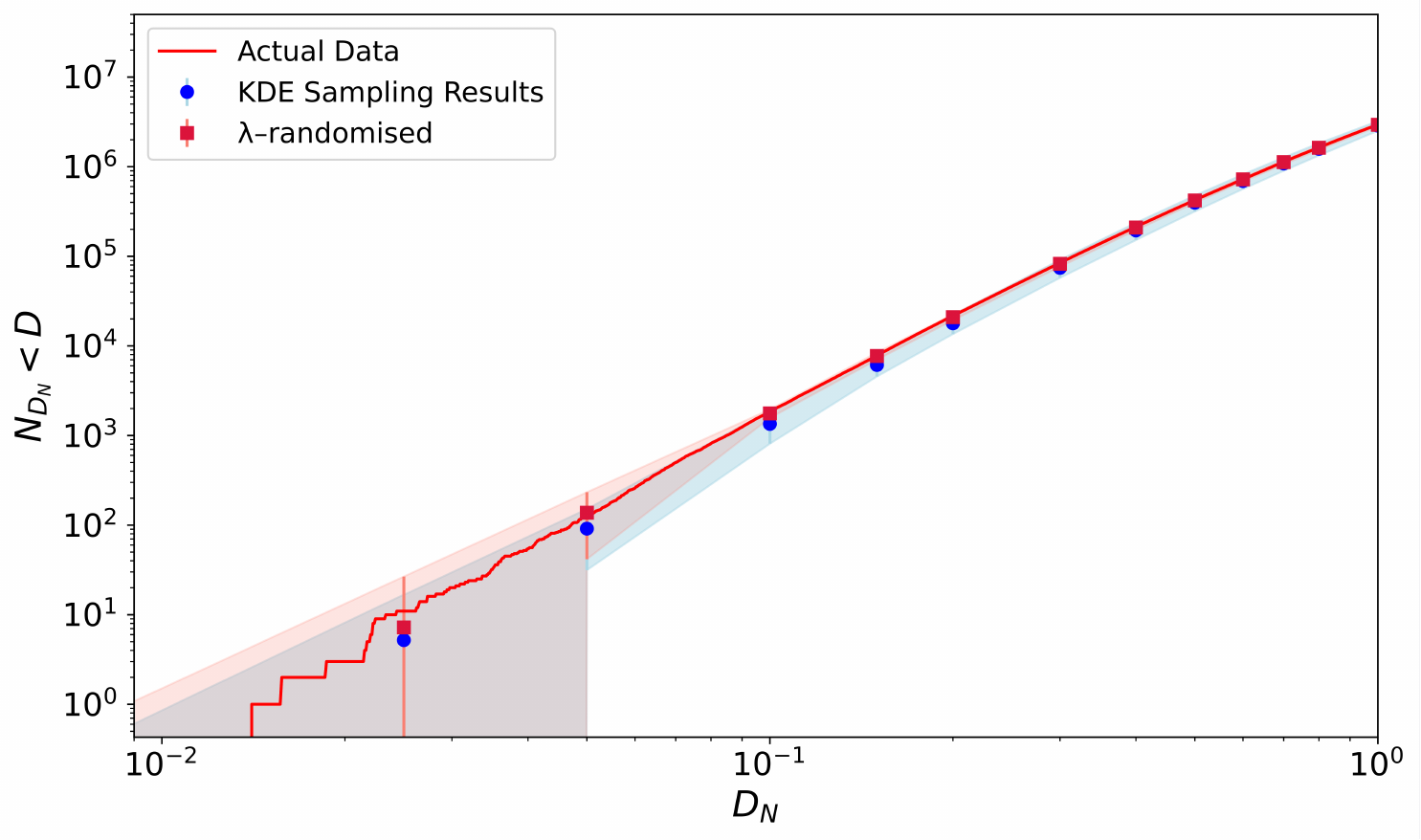}
        \caption{\centering  }
        \label{fig:result4}
    \end{subfigure}

    \vspace{0.4cm}

    \begin{subfigure}{0.48\linewidth}
        \centering
        \includegraphics[width=\linewidth]{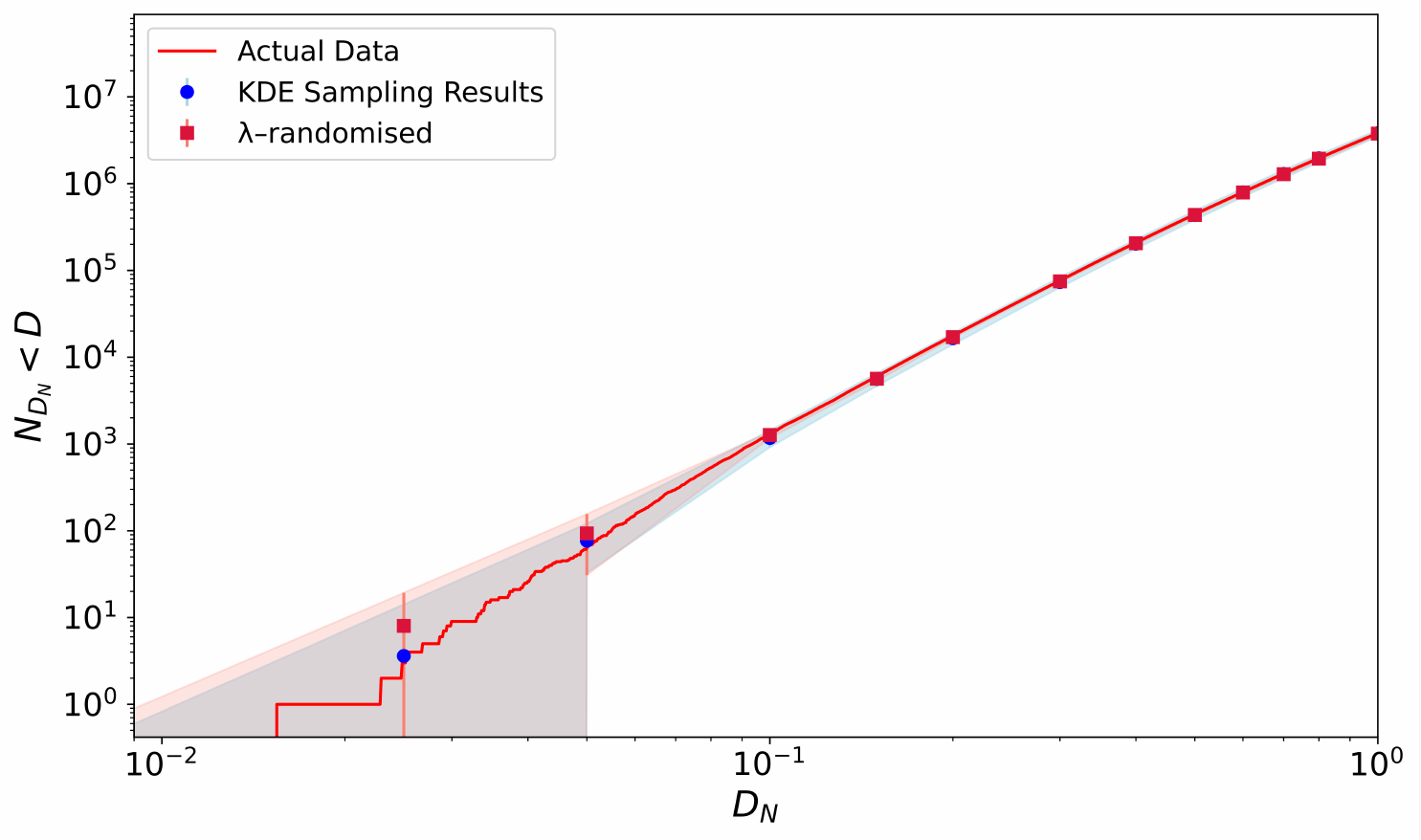}
        \caption{\centering  }
        \label{fig:result5}
    \end{subfigure}
    \hfill
    \begin{subfigure}{0.48\linewidth}
        \centering
        \includegraphics[width=\linewidth]{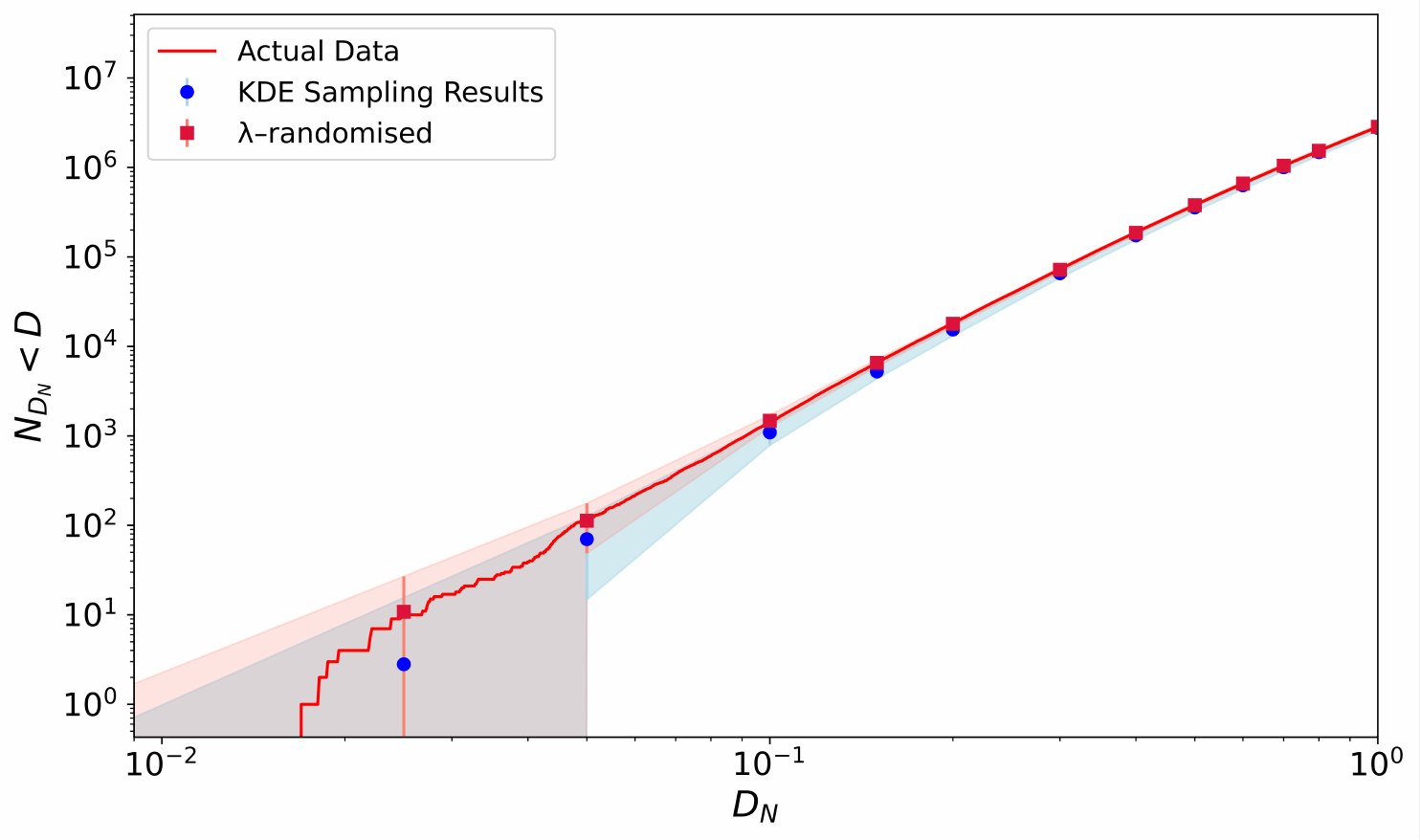}
        \caption{\centering  }
        \label{fig:result6}
    \end{subfigure}

    \caption{Same as Fig. 3 but for 535 potential meteorite falls $>$1\,g observed by GFO, FRIPON, or FRIPON sensors and NEA theoretical impact radiants.}
    \label{fig:535_CSDs}
\end{figure*}

Figures~\ref{fig:46_CSDs} and \ref{fig:535_CSDs} present the main test of this work: the CSDs of $D_N$ for (i) the 46 recovered falls and (ii) the 535 potential $>$1\,g falls, each compared against six independent NEA radiant catalogues (one per panel). 
In every panel, the red curve (labelled `Actual data') is the empirical cumulative count of all possible fall-radiant pairs with $D_N$ below a threshold. The two shaded regions represent the expected range of values for random associations using two different methods: the blue region is the $3\sigma$ range from Monte Carlo draws of the KDE-modelled sporadic background (Sect.~\ref{subsec:sporadic}), and the red region is the $3\sigma$ range from samples with uniformly randomised $\lambda_\odot$. 
The interpretation is straightforward: a statistically significant stream-like excess would manifest as the `actual data' curve deviating from a power law and rising above both the $3\sigma$ blue and red regions, particularly at low $D_N$. Small differences among panels are expected because each radiant method enforces Earth intersection via different element adjustments (Sect.~\ref{subsec:radiant_methods}).

\citet{carbognani2023identifying} argued that \mbox{$82\%$} of meteorite--NEA pairs with $D_{N}\!<\!0.06$ should be genuine. Using their NEODyS/ORBFIT radiants, we were able to reproduce all pairs listed in their Table 3. The observed CSD (red curve in Fig.~\ref{fig:46_CSDs}) sits above their black curve by a constant offset. This vertical shift is entirely explained by the additional $\sim4\,300$ Earth-crossing NEAs included here (\mbox{$20\,516$} versus\ \mbox{$16\,227$}); once normalised, the two CSDs coincide. The expected amount of stochastic similarity between the NEODyS/ORBFIT radiants and 46 recovered meteorites is estimated by (i) randomising only $\lambda_\odot$ as in \citet{carbognani2023identifying} (red region) and (ii) the KDE synthetic sporadic background introduced in Sect.~\ref{subsec:sporadic} (blue region). The observed CSD never exceeds the $3\sigma$ expectation envelope (Figs.~\ref{fig:46_CSDs}a–f), i.e. the amount of similarity between the NEA population and the 46 recovered falls is completely consistent with levels expected due to random associations. The single, marginal excursion for method~H at $D_{N}\!\simeq\!0.02$ remains well within the spread imposed by fireball radiant-velocity uncertainties and therefore is not significant.

Additionally, for random points in a $d$-dimensional space, the theoretical CSD follows a power law where $N(<\!D)\propto D^{d}$; asteroid-pair studies using five osculating elements indeed find slopes $\alpha\!\approx\!4.7$ \citep{vokrouhlicky2008pairs}. $D_{N}$ is built from four geocentric parameters, and thus the expected index drops to $\alpha\!\approx\!4$. A linear least-squares fit to our observed CSDs across the six radiant catalogues yields $\alpha=3.75$–$3.88$ (except Method~Q, where $\alpha\simeq4.35$) with uncertainties of $0.21$–$0.30$, all statistically consistent with $\alpha_{\rm th}=4$. These uncertainties were estimated by a 2{,}000-sample parametric bootstrap of the per-bin increments and by taking half the spread of slopes when varying the upper fit bound \(D_N^{\max}\in[0.08,0.12]\); the quoted \(\sigma\) combines the two in quadrature. The slightly shallower slope obtained from a broader fit (our nominal $\alpha\simeq3.8$) likely reflects finite-range effects and mild metric non-linearity in $D_{N}$. In contrast, the grey `random' line plotted by \citet{carbognani2023identifying} is an order of magnitude too low and, by construction (because it represents random associations), should be strictly linear; implying potentially a numerical error rather than a physical signal. Additionally, the grey `random' curve is only one sampling by \citet{carbognani2023identifying}, and does not encapsulate the confidence region but rather just one single estimate, which can vary significantly as we do not know the true sporadic population, especially when only based on 38 falls. 

Similarly, as seen in Fig.~\ref{fig:535_CSDs}, the CSD analysis for the 535 fireballs that satisfy the $m\!\geq1$\,g and 20\% deceleration criterion, does not show evidence of statistically significant clustering at low $D_{N}$ values. Every red curve sits comfortably inside the blue (KDE) and red (random–$\lambda_\odot$) $3\sigma$ regions. Thus, even when the meteorite-dropping sample is an order of magnitude larger, there is still no clear evidence of recently formed meteorite-dropping streams. 

A mild systematic offset appears at \mbox{$D_{N}\!\sim\!0.1$} in the NEODyS/ORBFIT panel (Fig.~\ref{fig:535_CSDs}a), where the observed CSD rises above the KDE envelope but not the randomised $\lambda_\odot$. The same minor feature is present, though entirely contained within $3\sigma$, for the analogous W-method based on 34\,836 NEAs. In no case does the observed CSD fall outside both $3\sigma$ regions; thus, this is likely a shortcoming of the KDE smoothing in this instance. In short, none of the six catalogues shows the low–$D_{N}$ surplus required for a detection, defined here as the `Actual data' curve rising above the $3\sigma$ envelopes of both null backgrounds at small $D_{N}$. This non-detection implies that any detectable fall–NEA stream component is at most of order $\sim10^{-3}$ of all falls (i.e. a larger contribution would have produced a visible excess). This is a detectability limit for our CSD test, not a formal binomial bound on the underlying fraction; for reference, a one–sided $3\sigma$ upper limit with zero detections in $N\!=\!535$ falls is $\lesssim0.5\%$. Streams of the order of the  $\sim10^{-3}$ level may exist but would be below the sensitivity of the present analysis.

\section{Discussion}\label{sec:discussion}

Analysis of 46 recovered meteorites and 535 possible $\geq$1\,g meteorite-dropping fireballs shows that the CSDs all follow an almost perfect power law at small $D_{N}$ values, enclosed by the $3\sigma$ envelopes expected for random pairings (Figs.~\ref{fig:46_CSDs} and \ref{fig:535_CSDs}). In other words, the amount of similarity found between the NEA geocentric radiants and the meteorite-dropping fireballs is fully consistent with that expected due to random associations. There are two instances where the observed CSDs step outside a single $3\sigma$ envelope (Figs.~\ref{fig:46_CSDs}d and~\ref{fig:535_CSDs}a), but neither satisfies our detection criterion (a low–$D_N$ surplus above both null envelopes). In the 46–fall case, the excursion occurs only against the $\lambda_\odot$–randomised null and disappears once the reported meteor radiant and velocity uncertainties are propagated through the CSD, as those uncertainties smear low-$D_N$ pairs and reduce any apparent excess. In the 535–fall NEODyS/ORBFIT panel, the modest rise above the KDE envelope occurs near $D_N\!\sim\!0.1$, is not seen in the other five radiant catalogues (including the similar W method), and is therefore most likely a local KDE-smoothing artefact rather than a coherent stream signal. Taken together, these checks show no low-$D_N$ surplus consistent with a recently formed, detectable stream.

There are two instances where the observed CSDs step outside a single $3\sigma$ envelope (Fig.~\ref{fig:46_CSDs}d; Fig.~\ref{fig:535_CSDs}a), but neither satisfies our detection criterion (a low-$D_N$ surplus above both null envelopes). In the 46–fall case, the excursion occurs only against the $\lambda_\odot$–randomised null and disappears once the reported meteor radiant/velocity uncertainties are propagated through the CSD, as those uncertainties smear low–$D_N$ pairs and reduce any apparent excess. In the 535–fall NEODyS/ORBFIT panel, the modest rise above the KDE envelope occurs near $D_N\!\sim\!0.1$, is not seen in the other five radiant catalogues (including the similar W–method), and is therefore most likely a local KDE-smoothing artefact rather than a coherent stream signal. Taken together, these checks show no low–$D_N$ surplus consistent with a recently formed, detectable stream.

From the widths of the $3\sigma$ envelopes at the smallest accessible separations, we inferred an order-of-magnitude detectability limit: any stream component contributing $\gtrsim 10^{-2}$ of all falls would produce a low–$D_{N}$ surplus that rises above null regions; since no such excess is seen in any catalogue, the stream fraction is $\lesssim 10^{-3}$ if present. This limit is consistent with \citet{shober2025decoherence}, who, using the classical \(D_{SH}\), \(D^{\prime}\), and \(D_{H}\) orbital similarity criteria applied to osculating elements, found no statistically significant meteorite streams amongst meteorite dropping fireballs. The present analysis extends that test through an entire secular cycle for every NEA radiant and still uncovers no hidden concentrations. Even if every tidally generated NEA cluster contributed meteorite-sized fragments, the stream fraction would rise only to the \(\simeq0.2\%\), an upper bound estimated by \citet{shober2025perihelion} presuming complete overlap between the identified NEA clusters of \citet{shober2025decoherence} and the 535 potential fall dataset.  

It is important to note that, within clustering studies like this one, the implicit assumption is that the broad features observed in our sporadic distribution are not related to streams. If someone would like to make the argument that, still, many meteorites are related to NEAs and that these broad features are actually indicative of a recent meteoroid stream activity, they would have to show that the orbital distribution expected for larger meteoroids in near-Earth space is not consistent with the orbital distribution expected purely based on the relative abundances coming from different source asteroid families in the main belt. Recent work indicates the opposite: \citet{broz2024young} found that most meteorites in our collections today originate from younger asteroid families, and their meteoroid transfer model called METEOMOD\footnote{\url{https://sirrah.troja.mff.cuni.cz/~mira/meteomod/}} reproduces the principal features of the observed FRIPON fireball velocity distribution (see Fig.~12 in \citealp{broz2024source}). Thus, the main, large-scale features of the fall population can be explained without invoking a substantial contribution from detectable streams of recently released material. Differences that remain are consistent with physical processing (e.g. thermal fragmentation during low-perihelion phases) rather than with a dominant stream component \citep{shober2025perihelion}.
 
There is significant evidence to suggest recent modification and activity within the near-Earth meteoroid and asteroid populations  \citep{borovivcka2015small,granvik2016super,lauretta2019episodes,bottke2020meteoroid,wiegert2020supercatastrophic,turner2021carbonaceous,scott2021short,shober_carbonaceous,granvik2024tidal,shober2025perihelion,jenniskens2025review}. However, this activity does not form large meteorite-dropping streams (\citealp{shober2025decoherence,chow2025decameter}, and this work). Recent activity can generally modify the orbital distribution of meteoroids, but direct identification of immediate precursor asteroids for meteorites seems unlikely and would necessitate other compositional or physical constraints in addition to dynamical justifications to make a convincing argument in the future. 

\section{Conclusions}
\label{sec:conclusions}

We investigated whether instrumentally observed meteorite falls or meteorite-dropping fireballs form statistically significant streams with NEAs when similarity is evaluated in
four-dimensional geocentric space \((U,\theta,\phi$,and $\lambda_\odot)\) using the $D_N$ criterion \citep{valsecchi1999meteoroid}. We compared 46 recovered meteorites and 535 candidate meteorite-dropping fireballs to six theoretical NEA geocentric radiant catalogues, based on $>30\,000$ NEOs. For each pairing, we built CSDs and benchmarked them against $10^7$–$10^8$ Monte Carlo trials drawn from (i) a uniform $\lambda_\odot$ distribution and (ii) kernel density estimates of the observed sporadic background.
 Our main results are:

 \smallskip
  1. None of the six impact radiant catalogues exhibits the low-$D_{N}$ surplus required for a detection, i.e. the observed cumulative similarity curve never rises above the $3\sigma$ envelopes of both null backgrounds at small $D_{N}$. All 46–fall CSDs lie entirely within the expected bands, and the 535 potential fall set shows only a single, likely method-specific fluctuation near $D_{N}\!\sim\!0.1$ that is absent in the other catalogues and remains inside the alternate null’s $3\sigma$ envelope. We therefore find no evidence of meteorite–NEA streams. The fitted small-$D_{N}$ slope, $\alpha=3.8\pm0.25$, is also consistent with the $N(<\!D)\propto D^{4}$ behaviour expected for a stochastic four–dimensional population.

\smallskip
  2. The upper limit on any hidden stream component is, therefore, at most $\sim0.1\%$ of all known meteorite falls. Extending the search through the full secular precession cycle of each NEA does not reveal additional pairings.

\smallskip
  3. These findings do not rule out recent physical activity (collisions, tidal disruption, thermal fragmentation, etc.) in the NEA population. Such processes are well documented, but studies show that this activity does not generate coherent, meteorite-dropping streams.

\medskip
\noindent In short, geocentric-parameter clustering provides no evidence that a substantial fraction of meteorites arrive from contemporary NEA streams. Any future attempt to link individual meteorites to immediate-precursor NEAs will require considerable additional constraints beyond simple orbital similarity.

\begin{acknowledgements}
This project has received funding from the European Union’s Horizon 2020 research and innovation programme under the Marie Skłodowska-Curie grant agreement No945298 ParisRegionFP. 

The Global Fireball Observatory and data pipeline is enabled by the support of the Australian Research Council (DP230100301, LE170100106).

FRIPON was initiated by funding from ANR (grant N.13-BS05-0009-03), carried by the Paris Observatory, Muséum National d’Histoire Naturelle, Paris-Saclay University and Institut Pythéas (LAM-CEREGE). VigieCiel was part of the 65 Millions d’Observateurs project, carried by the Muséum National d’Histoire Naturelle and funded by the French Investissements d’Avenir program. FRIPON data are hosted and processed at Institut Pythéas SIP (Service Informatique Pythéas), and a mirror is hosted at IMCCE (Institut de MécaniqueCéleste et de Calcul des Éphémérides / Paris Observatory). 

This research used Astropy, a community-developed core Python package for Astronomy \citep{robitaille2013astropy}.

\end{acknowledgements}

\bibliographystyle{aa} 
\bibliography{references}

\begin{appendix}
\section{List of the 46 recovered meteorite falls}
\label{app:meteorite_table}
\begin{table}[ht!]
  \centering
  \caption{Instrumentally observed recovered meteorite falls used in this study.}
  \label{tab:meteorite_falls}

  \tiny

  \resizebox{\textwidth}{!}{
    \begin{tabular}{@{} 
      l  
      l   
      l   
      l   
      r   
      r   
      l   
    @{}}
      \hline\hline
Name & Type & Location & Date & $m_{\infty}$ (kg) & $m_{found}$ (kg) & Source \\
\hline
Pribram & H5 & Czech Rep. & 1959/04/07 & $<$5000 & 5.56 & \citet{ceplecha1961multiple} \\ 
Lost City & H5 & USA & 1970/01/04 & 165 & 17 & \citet{mccrosky1971lost} \\ 
Innisfree & L5 & Canada & 1977/02/06 & 42 & 4.58 & \citet{halliday1978innisfree} \\ 
Benešov & LL3.5, H5 & Czech Rep. & 1991/05/07 & 4100 & 0.01 & \citet{spurny2014reanalysis} \\ 
Peekskill & H6 & USA & 1992/10/09 & $<$10\,000 & 12.57 & \citet{brown1994orbit} \\ 
Tagish Lake & C2-ung. & Canada & 2000/01/18 & 75000 & 10 & \citet{brown2000fall} \\ 
Moravka & H5 & Czech Rep. & 2000/05/06 & 1500 & 0.63 & \citet{borovicka2003moravka} \\ 
Neuschwanstein & EL6 & Germany & 2002/04/06 & 300 & 6.19 & \citet{spurny2003photographic} \\ 
Park Forest & L5 & USA & 2003/03/27 & 11000 & 18 & \citet{brown2004orbit} \\ 
Villalbeto de la Peña & L6 & Spain & 2004/01/04 & 760 & 3.5 & \citet{llorca2005villalbeto} \\ 
Bunburra Rockhole & Eurcite & Australia & 2007/07/20 & 22 & 0.32 & \citet{bland2009anomalous} \\ 
Almahata Sitta & Ureilite +other & Sudan & 2008/10/07 & 83000 & 3.95 & \citet{jenniskens2009impact} \\ 
Buzzard Coulee & H4 & Canada & 2008/11/21 & 15\,000 & 41 & \citet{fry2013physical} \\ 
Maribo & CM2 & Denmark & 2009/01/17 & 1500 & 0.03 & \citet{borovivcka2019maribo} \\ 
Jesenice & L6 & Slovenia & 2009/04/09 & 170 & 3.67 & \citet{spurny2010analysis} \\ 
Grimsby & H5 & Canada & 2009/09/26 & 33 & 0.22 & \citet{brown2011fall} \\ 
Kosice & H5 & Slovakia & 2010/02/28 & 3500 & 4.3 & \citet{borovivcka2013kovsice} \\ 
Mason Gully & H5 & Australia & 2010/04/13 & 40 & 0.02 & \citet{dyl2016characterization} \\ 
Križevci & H6 & Croatia & 2011/02/04 & 25-100 & 0.29 & \citet{borovivcka2015instrumentally} \\ 
Sutter's Mill & C, CM2 & USA & 2012/04/22 & 50000 & 0.99 & \citet{jenniskens2012radar} \\ 
Novato & L6 & USA & 2012/10/18 & 80 & 0.31 & \citet{jenniskens2014fall} \\ 
Chelyabinsk & LL5 & Russia & 2013/02/15 & 1\,200\,000 & $\sim$1000 & \citet{popova2013chelyabinsk} \\ 
Annama & H5 & Russia & 2014/04/18 & 472 & 0.17 & \citet{kohout2017annama} \\ 
Žďár nad Sázavou & L3 & Czech Rep. & 2014/12/09 & 170 & 0.05 & \citet{spurny2020vzvdar} \\ 
Porangaba & L4 & Brazil & 2015/01/09 & 0.976 & 0.976 & \citet{ferus2020elemental} \\ 
Sariçiçek & Howardite & Turkey & 2015/09/02 & 6000 - 20000 & 15.24 & \citet{unsalan2019saricciccek} \\ 
Creston & L6 & USA & 2015/10/24 & 10-100 & 0.688 & \citet{jenniskens2019creston} \\ 
Murrili & H5 & Australia & 2015/10/27 & 37.9$\pm$2.3 & 1.68 & \citet{sansom2020murrili} \\ 
Ejby & H5/6 & Denmark & 2016/02/06 & 250 & 8.94 & \citet{spurny2017atmospheric} \\ 
Dishchii'bikoh & LL7 & USA & 2016/06/02 & 3000-15000 & 0.0795 & \citet{jenniskens2020orbit} \\ 
Dingle Dell & L/LL5 & Australia & 2016/10/31 & 40 & 1.15 & \citet{devillepoix2018dingle} \\ 
Hamburg & H4 & USA & 2018/01/17 & 60 - 225 & 1 & \citet{brown2019hamburg} \\ 
Motopi Pan & Howardite & Botswana & 2018/06/02 & 5500 & 0.214 & \citet{jenniskens2021impact} \\ 
Ozerki & L6 & Russia & 2018/06/21 & 94000$\pm$20000 & 6.5 & \citet{kartashova2020investigation} \\ 
Arpu Kuilpu & H5 & Australia & 2019/06/01 & 0.91 & 0.031 & \citet{shober2022arpu} \\ 
Flensburg & C1-ungr. & Germany & 2019/09/12 & 10000-20000 & 0.0245 & \citet{borovička2021trajectory} \\ 
Cavezzo & L5-an & Italy & 2020/01/01 & 3.5 & 0.055 & \citet{gardiol2021cavezzo} \\ 
Novo Mesto & L5 & Slovenia & 2020/02/28 & 470 & 0.72 & \citet{vida2021novo} \\ 
Madura Cave & L5 & Australia & 2020/06/19 & 30-60 & 1.072 & \citet{Devillepoix_Madura_Cave} \\ 
Narashino & H5 & Japan & 2020/07/02 & ~ & 0.35 & \href{https://www.lpi.usra.edu/meteor/metbull.php?table\&code=72653}{Metbull} \\
Traspena & L5 & Spain & 2021/01/18 & 2620 & 0.527 & \citet{andrade2023traspena} \\ 
Winchcombe & CM2 & UK & 2021/02/28 & 12.5 & 0.602 & \citet{mcmullan2024winchcombe} \\ 
Antonin & L5 & Poland & 2021/07/15 & 50-500 & 0.35 & \citet{shrbeny2022analysis} \\ 
Golden & L/LL5 & Canada & 2021/10/04 & 70 & 2.2 & \citet{brown2023golden} \\ 
Saint-Pierre-le-Viger & L5-6 & France & 2023/02/13 & 780 & 1.2 & \citet{egal2025CX1} \\ 
Ribbeck & Aubrite & Germany & 2024/01/24 & 140 & 1.8 & \citet{spurny2024atmospheric} \\ 
      \hline
    \end{tabular}
  }
\end{table}

\end{appendix}

\end{document}